\documentclass{jpp}

\usepackage[utf8]{inputenc}
\usepackage[T1]{fontenc}

\usepackage{amsmath}
\usepackage{amsfonts}
\usepackage{amssymb}
\usepackage{textcomp}
\newcommand\redout{\bgroup\markoverwith
{\textcolor{red}{\rule[.5ex]{2pt}{0.4pt}}}\ULon}

\usepackage{url}
\usepackage{graphicx}
\usepackage[hidelinks]{hyperref}
\usepackage[capitalize]{cleveref}
\usepackage{multirow}
\usepackage{interval}
\usepackage{mdframed}
\usepackage{soul}
\usepackage{xcolor}
\sethlcolor{yellow}

\newmdenv[linecolor=green,backgroundcolor=green]{highlighted}

\usepackage[symbol]{footmisc}

\shorttitle{Using Deep Learning to Design High Aspect Ratio Fusion Devices}
\shortauthor{P. Curvo, D. R. Ferreira, R. Jorge}

\title{Using Deep Learning to Design High Aspect Ratio Fusion Devices}

\author{P. Curvo\aff{1}\corresp{\email{pedro.curvo@tecnico.ulisboa.pt}},
        D. R. Ferreira\aff{1},
        R. Jorge\aff{2}}

\affiliation{\aff{1}Instituto de Plasmas e Fusão Nuclear, Instituto Superior Técnico, Universidade de Lisboa, 1049-001 Lisbon, Portugal
\aff{2}Department of Physics, University of Wisconsin-Madison, Madison, Wisconsin 53706, USA}

\begin{document}

\maketitle

\begin{abstract}
The design of fusion devices is typically based on computationally expensive simulations. This can be alleviated using high aspect ratio models that employ a reduced number of free parameters, especially in the case of stellarator optimization where non-axisymmetric magnetic fields with a large parameter space are optimized to satisfy certain performance criteria. However, optimization is still required to find configurations with properties such as low elongation, high rotational transform, finite beta, and good fast particle confinement. In this work, we train a machine learning model to construct configurations with favorable confinement properties by finding a solution to the inverse design problem, that is, obtaining a set of model input parameters for given desired properties. Since the solution of the inverse problem is non-unique, a probabilistic approach, based on mixture density networks, is used. It is shown that optimized configurations can be generated reliably using this method.
\end{abstract}

\section{Introduction}

\begin{figure} 
    \centering
    \includegraphics[width=0.4\textwidth]{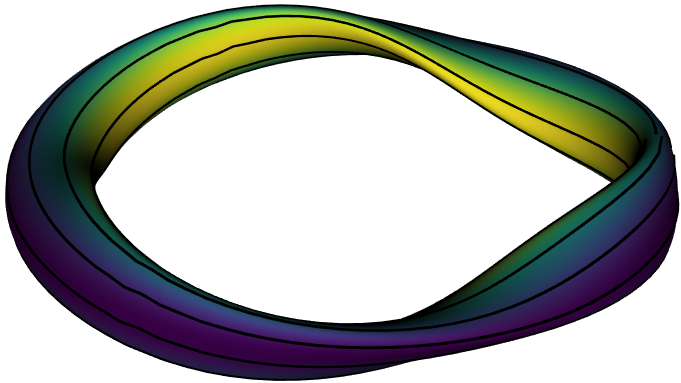}
    \caption{Plasma boundary of a quasisymmetric stellarator with three field periods, $n_{\text{fp}}=3$. The colors represent the magnetic field strength at the boundary and a magnetic field line is shown in black.}
    \label{fig:stellarator}
\end{figure}

Stellarators are a type of magnetic confinement fusion device that have toroidal geometry and are non-axisymmetric (see \cref{fig:stellarator}). Stellarators are inherently current-free, enabling steady-state plasma operation. Because of this, they are one of the leading candidates for future fusion energy power plants~\citep{Boozer_2020}. In these devices, the magnetic field is twisted by a rotation of the poloidal cross-section of stretched flux surfaces around the torus, and by making the magnetic axis non-planar~\citep{Spitzer1958, Mercier1964}. Stellarators are inherently current-free, enabling steady-state plasma operation. Because of this, they are one of the leading candidates for future fusion energy power plants~\citep{Boozer_2020}. Due to their complex geometries, stellarators may present difficulties in confining charged particles, especially alpha particles resulting from fusion reactions~\citep{Helander2014}. Therefore, they need accurately shaped magnetic fields to confine trapped particles effectively. To achieve this, their configurations are usually optimized using numerical methods. However, the optimization process is complex due to the high-dimensional space of plasma shapes, which includes numerous local minima~\citep{Bader2019}. While local optimization algorithms can find specific configurations, they do not offer a global view of the solution space. The high dimensionality makes global optimization challenging and renders comprehensive parameter scans impractical~\citep{Landreman_2022}.

To address these challenges, a near-axis method is commonly employed~\citep{Garren1991a, Landreman2021, Landreman2020a}. This method makes use of an approximate magnetohydrodynamic (MHD) equilibrium model by expanding in powers of the distance to the axis, leading to a small set of one-dimensional ordinary differential equations~\citep{Landreman_2022}, therefore reducing the computational costs significantly~\citep{Mercier1964, Solovev1995, Garren1991}. As a result, physical intuition can be more easily obtained, and high-resolution multidimensional parameter scans become more feasible, facilitating the generation of extensive databases of stellarator configurations.

In this work, we use the near-axis expansion to second order to generate configurations with finite plasma $\beta=2 \mu_0 p/B^2$ where $p$ is the plasma pressure and $B$ is the magnetic field strength. This allows us to find Mercier stable configurations by selecting devices that satisfy the Mercier criterion, $D_\text{Merc}>0$~\citep{Landreman2020a}. Such configurations have a positive magnetic well and are robust against certain MHD instabilities. In addition to MHD stability, we will also target configurations with low aspect ratio, small elongation, large rotational transform, and quasisymmetry, i.e., good particle confinement \citep{Paul_2022}.
Such quantities can be computed using already available software packages such as \texttt{pyQSC},\footnote[2]{\url{https://github.com/landreman/pyQSC}} which receives a set of design parameters (such as axis shape) and computes a set of properties (such as level of quasisymmetry). However, not all configurations are desirable. For most input parameters, the resulting configuration may be unacceptable due to factors such as a too-small volume of plasma, varying levels of quasisymmetry, low rotational transform, or overly large elongation. Therefore, it is essential to verify whether the configurations meet specific criteria. This verification can be time-consuming and often requires running the near-axis method multiple times to achieve a viable configuration or resort to numerical optimization. This prompts the question of whether it is possible to perform inverse design, i.e., to determine the input parameters from a given set of desired properties, hence creating a more convenient and efficient method for generating optimized stellarator configurations. This is the main goal of this work.

Since analytically inverting the equations in the near-axis method is not feasible due to their differential integral character, a practical solution involves employing a machine learning model, specifically a neural network as a universal approximator~\citep{Hornik1989} to tackle the inverse problem. By training on a dataset of near-axis configurations, the neural network can learn either a forward mapping, from design parameters to configuration properties, or an inverse mapping, from configuration properties to design parameters. However, this inverse problem is ill-posed as multiple sets of design parameters can yield the same configuration properties {(this was also observed in the database used in this work)}. {This means that the standard stellarator design formulation is not bijective, in that it lacks a unique, one-to-one correspondence between design parameters and configuration properties.}
{As with other inverse design problems, using a neural network in this context can result in predictions that represent an average of multiple possible design parameter sets for given configuration properties, rather than a specific solution. This averaging effect, described by~\cite{Bishop1994}, can lead to inaccurate outcomes, as the network generalizes over multiple valid solutions instead of a unique parameter set.}
To overcome this challenge, we approximate the probability distribution of the design parameters conditioned on the configuration properties. This distribution, which can be multimodal~\citep{McLachlan1988}, allows us to sample design parameters based on the desired configuration properties. To achieve this, a probabilistic machine learning model, namely the Mixture Density Networks (MDNs) model~\citep{Bishop1994}, is used to solve the inverse problem of stellarator optimization, together with the near-axis expansion method.\footnote[3]{The code developed during this work is available at
\url{https://github.com/pedrocurvo/MLStellaratorDesign}}

\section{Physical Model}

In this section, we describe the near-axis expansion method used to find quasisymmetric stellarators. Quasisymmetry is an effective strategy for confining trapped particles~\citep{Helander2014, Nuhrenberg1988} and consists of a continuous symmetry of the magnitude $B$ of the magnetic field  \textbf{B} that yields a conserved quantity and enhances particle confinement. Near the magnetic axis, two types of quasisymmetry are possible, namely quasi-axisymmetry (QA), where \( B = B(r, \theta) \) and quasi-helical symmetry (QH), where \( B = B(r, \theta - N \varphi) \).
Here, \( (\theta, \varphi) \) are the Boozer poloidal and toroidal angles~\citep{Boozer1981}, \( N \) is an integer, \( r \) is defined as $r = \sqrt{2\psi/B_0}$ where $\psi$ represents the magnetic toroidal flux and acts as a radial coordinate and $B_0$ is the magnetic field strength on the magnetic axis.

The method for generating stellarator configurations in a near-axis expansion complements traditional stellarator optimization, which typically involves parameterizing the boundary shape of a finite aspect ratio plasma and using a 3D MHD equilibrium code to evaluate the objective function.
{The magnetic field equilibrium and plasma pressure are related via the ideal MHD equation $\mathbf J \times \mathbf B = \nabla p$ with $\mathbf J=\nabla \times \mathbf B/\mu_0$ the plasma current.}
Instead, in the near-axis method, we parameterize the axis curve and find the Taylor series coefficients of $\textbf{B}$ in powers of $r$ that allow for quasisymmetry~\citep{Garren1991, Landreman2019b, Jorge2020}. While the near-axis method is necessarily approximate, it is orders of magnitude faster than standard methods, with a reduced parameter space, therefore allowing for broader parameter scans. Ultimately combining both approaches can be advantageous: the near-axis method can identify viable configurations, which can then be refined through conventional optimization.

In the near-axis expansion method, the magnetic axis \( \mathbf{r}_0 = R(\phi) \mathbf{e}_R + Z(\phi) \mathbf{e}_z\) is typically represented in cylindrical coordinates \( (R, Z, \phi) \) using a finite Fourier series,
\begin{equation}
    \begin{aligned}
        R(\phi) & = \sum_{n=0}^{N_F} R_{cn} \cos(n_{\text{fp}} n \phi),
        & Z(\phi) & = \sum_{n=1}^{N_F} Z_{sn} \sin(n_{\text{fp}} n \phi),
    \end{aligned}
    \label{eq:RZ}
\end{equation}
where \( n_{\text{fp}} \) is the number of field periods and a finite maximum Fourier number \( N_F \) is chosen. Stellarator symmetry is assumed. The remaining input parameters are the coefficients of the magnetic field strength
\begin{equation}
    \begin{split}
        B = & B_0 [1 + r \bar{\eta} \cos(\vartheta)] + r^2[B_{20}+B_{2c} \cos(2\vartheta)],
    \end{split}
    \label{eq:magnetic_B}
\end{equation}
namely $B_{0}$, $\bar{\eta}$ and $B_{2c}$, and the plasma pressure
\begin{equation}
    p=p_2 r^2,
    \label{eq:pressure}
\end{equation}
with $\vartheta = \theta-N \varphi$. Here, $B_0$ is chosen to be $1\,\text{T}$ and, following~\citet{Landreman2019b}, $B_{20}$ is taken to be a function of $\varphi$, with exact quasisymmetry corresponding to $B_{20}$ being a scalar constant. The total plasma current on-axis is taken to be $I_{2}=0$. Henceforth, the input parameter space for optimization consists of \( \{R_{cn}, Z_{sn}, n_{\text{fp}},\bar{\eta}, B_{2c}, p_2\} \), as described in~\cref{tab:input_parameters}. The output properties are presented in~\cref{tab:output_properties}. 
{The magnetic field equilibrium and plasma pressure are related via the ideal MHD equation $\mathbf J \times \mathbf B = \nabla p$ with $\mathbf J=\nabla \times \mathbf B/\mu_0$ the plasma current.}
{The proxy used for the maximum plasma radius is $r_\text{singularity}$ (see \citet{Landreman2021a}), and the proxy used for the plasma $\beta$ is the volume-averaged $\left<\beta\right>=-\mu_0 p_2 r_\text{singularity}^2/B_0$ (see \citet{Landreman_2022}).}
{The number of degrees of freedom used to optimize stellarator devices has then been reduced from typically $\sim$100 plasma boundary coefficients to $\sim10$ near-axis coefficients.}

\begin{table} 
    \centering
    \begin{tabular}{l|p{0.7\linewidth}}
        \textbf{Input} & \textbf{Description} \\
        \hline
        $R_{c1}$ & First Fourier coefficient of $R(\phi)$ in~\cref{eq:RZ}. \\
        $R_{c2}$ & Second Fourier coefficient of $R(\phi)$ in~\cref{eq:RZ}. \\
        $R_{c3}$ & Third Fourier coefficient of $R(\phi)$ in~\cref{eq:RZ}. \\
        $Z_{s1}$ & First Fourier coefficient of $R(\phi)$ in~\cref{eq:RZ}. \\
        $Z_{s2}$ & Second Fourier coefficient of $R(\phi)$ in~\cref{eq:RZ}. \\
        $Z_{s3}$ & Third Fourier coefficient of $R(\phi)$ in~\cref{eq:RZ}. \\
        $\bar{\eta}$ & First order Taylor series coefficient of $B$ in~\cref{eq:magnetic_B} \\
        $B_{2C}$ & Second order Taylor series coefficient of $B$ in~\cref{eq:magnetic_B}. \\
        $n_{\text{fp}}$ & Number of field periods of the device.\\
        $p_2$ & Second order Taylor series coefficient of $p$ in~\cref{eq:pressure}. \\
    \end{tabular}
    \caption{Input parameters for the near-axis model}
    \label{tab:input_parameters}
\end{table}

\begin{table} 
	\centering
	\begin{tabular}{l|p{0.7\linewidth}}
		\textbf{Output} & \textbf{Description} \\
		\hline
		$\text{axis length}$ & Length of the magnetic axis \\
		$\iota$ & Rotational transform on-axis. \\
		$\text{max elongation}$ & Ratio of the major to minor semi-axis cross-section. \\
		$\text{min } L_{\nabla B}$ & Scale length of the magnetic field gradient. \\
		$\text{min } R_0$ & Minimum of the radial coordinate $R$ of the axis. \\
		$r_{\text{singularity}}$ & Maximum allowed radial coordinate for the boundary.\\
		$L_{\nabla\nabla B}$ & Scale length of the magnetic field Hessian. \\
		$B_{20_{\text{variation}}}$ & Degree of quasisymmetry. \\
		$\beta$ & {Volume-averaged plasma beta} $\left<\beta\right>=-\mu_0 p_2 r_{\text{singularity}}^2 / B_0^2$. \\
		$D_{\text{Merc}}\times r^2$ & Lowest order Mercier criterion coefficient. \\
	\end{tabular}
    \caption{Output parameters from the near-axis model}
    \label{tab:output_properties}
\end{table}

Although the near-axis expansion substantially reduces the number of free parameters, the optimization process may be computationally expensive, depending on the target parameters. Furthermore, it is necessary to compute the mapping between Boozer and Cartesian coordinates for a given surface to identify if the configuration possesses self-intersecting surfaces, making the process both time-consuming and resource-intensive. For this work, a viable configuration is one that meets the specific criteria outlined in~\cref{tab:good_stellarators}. Such parameters are similar to the ones outlined in~\citet{Landreman_2022}. {Those parameters also benefit the overall stability of the stellarators, e.g., $L_{\nabla B}$ is positively correlated with the coil-to-plasma distance, as demonstrated by~\citet{Kappel2024}, hence by constraining to larger values, we can obtain solutions with improved stability.} Informally, we will refer to stellarators that meet these criteria as \emph{good} stellarators, and those that do not as \emph{bad} stellarators.

\begin{table} 
    \centering
    \begin{tabular}{l|l}
        \textbf{Output Property} & \textbf{Range} \\
        \hline
        $\text{axis length}$ & $> 0.0$ \\
        $|\iota|$ & $\geq 0.2$ \\
        $\text{max elongation}$ & $\leq 10.0$ \\
        $\text{min }L_{\nabla B}$ & $\geq 0.1$ \\
        $\text{min }R_0$ & $\geq 0.3$ \\
        $r_{\text{singularity}}$ & $\geq 0.05$ \\
        $L_{\nabla\nabla B}$ & $\geq 0.1$ \\
        $B_{20_{\text{variation}}}$ & $\leq 5.0$ \\
        $\beta$ & $\geq 10^{-4}$ \\
        $D_{\text{Merc}}\times r^2$ & $> 0.0$ \\
    \end{tabular}
    \caption{Criteria for good stellarators with the major radius fixed at $R_{c0} = 1$m and magnetic field on-axis of $B_{0} = 1$T}
    \label{tab:good_stellarators}
\end{table}

\section{Mixture Models and Density Networks}
\label{section3c}

When dealing with non-unique inverse problems, we often encounter situations where there are multiple possible solutions for a given input. To effectively address these problems, it is essential to have a statistical distribution over the possible solutions rather than a single deterministic answer. The normal distribution is a common way to construct probability distributions, but in cases with multiple solutions, we require a multi-modal distribution, which can be achieved through a mixture model~\citep{McLachlan1988}. This model provides concentrated probabilities at various points, representing the different solutions. In this section, we describe the probabilistic models used in this work, namely mixture models, Gaussian models, and multivariate Gaussian mixtures.

A mixture model~\citep{McLachlan1988} is a statistical tool used to describe a population comprised of multiple subgroups without prior knowledge of individual data point memberships. It constructs a combined probability distribution for the entire population by integrating the probability distributions of each subgroup. Mixture models enable us to understand the characteristics of these subgroups using data from the entire population, even when the subgroup for each data point is unknown. These models are typically applied in clustering tasks, where data points are grouped into clusters, and density estimation, which involves estimating the distribution of the data itself.

A typical finite-dimensional mixture model $p(y | \lambda)$ is a combination of simple distributions $p_i(y)$ that can be represented as follows
\begin{equation}
	\label{eq:mixture_model}
	p(y | \lambda) = \sum_{i=1}^{K} \pi_i p_i(y),
\end{equation}
where \( p_i \) is the \( i^{\text{th}} \) component distribution, \( \pi_i \) is the mixture weight of the \( i^{\text{th}} \) component, and \( K \) is the number of components in the mixture. The mixture weights are non-negative and sum to 1, i.e., \( 0 \leq \pi_i \leq 1 \) and \( \sum_i \pi_i = 1 \).

To better understand mixture models, we re-express the model in a hierarchical framework. This involves introducing a latent variable \( z \in \{1, ..., K\} \) representing the component from which each data point is generated. This hierarchical approach not only provides a clear structure but also facilitates the inference process. Henceforth, each data point \( y \) is associated with a latent variable \( z \) that indicates the component it originates from. The prior distribution over the latent variables is governed by the parameters \( \pi = (\pi_1, ..., \pi_K) \), where \( \pi_i \) represents the probability that a data point belongs to component \( i \). Formally, we write
\[
p(z = k | \lambda) = \pi_k.
\]
Given that a data point \( y \) comes from component \( i \), it is generated according to a component-specific distribution \( p(y | \lambda_i) \). Thus, the conditional distribution of \( y \) given the latent variable \( z \) and the parameters \( \lambda \) is
\[
p(y | z = i, \lambda) = p_i(y) = p(y|\lambda_i).
\]
The complete set of parameters for this hierarchical model is \( \lambda = (\pi_1, ..., \pi_K, \lambda_1, ..., \lambda_K) \), where \( \pi \) represents the mixing proportions and \( \lambda_i \) represents the parameters specific to the \( i^{\text{th}} \) component.

The generative process for the data involves first selecting a specific component \( z \) and then drawing a sample \( y \) from the chosen component. By marginalizing over the latent variable, i.e., by summing over all possible states of $z$, we obtain the marginal distribution $p(y | \lambda)$ of the observed data
\begin{equation}
	p(y | \lambda) = \sum_{i=1}^{K} p(z = i | \lambda) p(y | z = i, \lambda) = \sum_{i=1}^{K} \pi_i p(y | \lambda_i).
\end{equation}
This formulation allows us to model complex, multi-modal data distributions effectively, capturing the diverse characteristics of the data through the combined influence of multiple simple components.

\begin{figure} 
    \centering
    \includegraphics[width=0.7\textwidth]{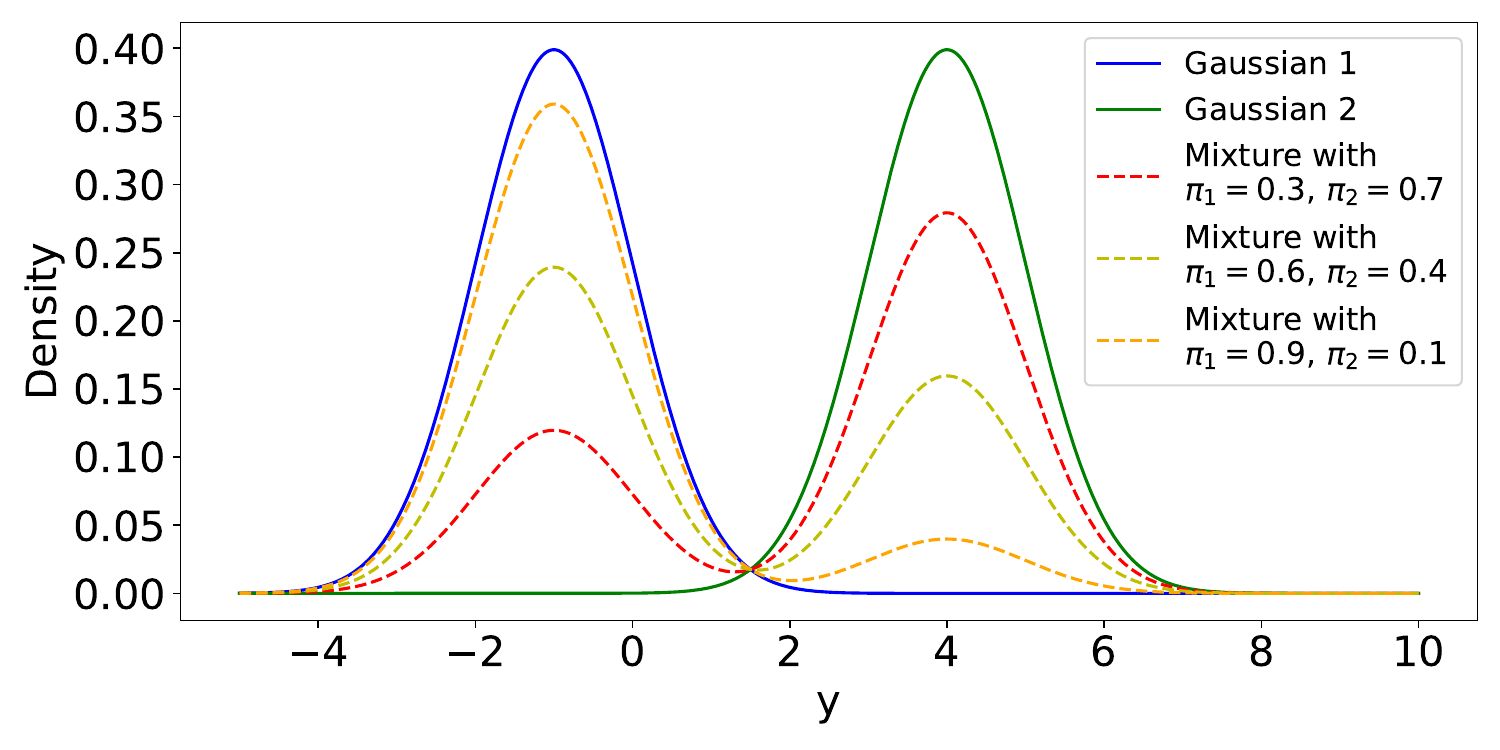}
    \caption{Example of mixture models with two components, each represented by a Gaussian distribution, illustrating how a mixture model forms from two distributions and the influence of mixture weights on data distribution modeling.}
    \label{fig:mixture_model}
\end{figure}

One of the most widely used mixture models, due to its simplicity and effectiveness in modeling complex data distributions, is the Gaussian Mixture Model (GMM), a specific type of mixture model where the component distributions are Gaussian distributions. The GMM is defined as
\begin{equation}
    \label{eq:GMM}
    p(y | \lambda) = \sum_{i=1}^{K} \pi_i \mathcal{N}(y | \mu_i, \sigma_i^2),
\end{equation}
where \( \mathcal{N}(y | \mu_i, \sigma_i^2) \) is a Gaussian distribution with mean \( \mu_i \) and variance \( \sigma_i^2 \), namely
\begin{equation}
    \label{eq:uni_gaussian}
    \mathcal{N}(y | \mu_i, \sigma_i^2) = \frac{1}{\sqrt{2\pi\sigma_i^2}} e^{ -\frac{(y - \mu_i)^2}{2\sigma_i^2} } .
\end{equation}
The GMM can approximate any continuous distribution to any arbitrary degree of accuracy by using a sufficient number of components~\citep{Goodfellow2016}. It is particularly useful for clustering and density estimation tasks, where the data distribution is complex and multi-modal. An example of a GMM with two components and different mixture weights is shown in~\cref{fig:mixture_model}.  
This figure illustrates how the GMM combines two Gaussian distributions with distinct means and variances, demonstrating three separate mixtures where each mixture is characterized by specific mixing coefficients, \( \pi_1 \) and \( \pi_2 \). These coefficients determine the relative influence of each Gaussian component in modeling the observed data distribution, showcasing the GMM's ability to represent complex data patterns through weighted combinations of simpler Gaussian distributions.

A generalization of the one-dimensional Gaussian distribution to multiple dimensions is called multivariate normal (MVN), also known as the multivariate Gaussian distribution. The MVN is one of the most widely used joint probability distributions for continuous random variables~\citep{KMurphy2023}. This popularity is due to its mathematical convenience and versatile applicability across a wide range of scenarios. Indeed, if we know the mean and variance of a dataset, but do not have other information such as, for example, skewness, kurtosis, domain-specific constraints, temporal dependencies, spatial correlations, or known outliers, the Gaussian distribution is the most \textit{unbiased} choice because it maximizes entropy under these constraints~\citep{Cover2012}.

The multivariate Gaussian distribution is defined as
\begin{equation}
	\mathcal{N} (y | \mu, \Sigma) = \frac{1}{(2\pi)^{D/2} |\Sigma|^{1/2}} e^{ -\frac{1}{2} (y - \mu)^T \Sigma^{-1} (y - \mu) },
\end{equation}
where \( y \) is a \( D \)-dimensional vector, \( \mu =\mathbb{E}[y] \) is the mean vector, and \( \Sigma = \text{Cov}[y] \) is the \( D \times D \) covariance matrix. The normalization constant is given by \( (2\pi)^{D/2} |\Sigma|^{1/2} \) to ensure that the distribution has a unit volume integral. 
The covariance matrices \( \Sigma \), in the context of multivariate Gaussian distributions, can be categorized into three groups. First, full covariance matrices are matrices with \( D(D + 1)/2 \) parameters, which are symmetric and positive definite, allowing them to capture existing correlations between variables.
Second, diagonal covariance matrices are matrices with \( D \) parameters, which are diagonal with zero off-diagonal elements. These matrices assume that the variables are independent of each other.
Lastly, there are spherical covariance matrices with one parameter, which are a scalar multiple of the identity matrix in the form \( \sigma^2 I_D \). These matrices assume that the variables have equal variance and are isotropic.

The full covariance matrix is the most general form of the multivariate Gaussian distribution, and it can represent existing correlations between variables. However, it is also the most computationally expensive, since it requires the inversion of a \( D \times D \) matrix. The diagonal covariance matrix, on the other hand, assumes that the variables are independent, and the spherical covariance matrix assumes that the variables are isotropic, both simplifying computation but potentially oversimplifying real-world correlations.

Using multiple MVNs as components in a Mixture Model results in what is known as the Multivariate Gaussian Mixture Model (MGMM). This model is a generalization of the GMM to the multivariate case and is defined as
\begin{equation}
	\label{eq:mgmm}
	p(y | \lambda) = \sum_{i=1}^{K} \pi_i \mathcal{N}(y | \mu_i, \Sigma_i),
\end{equation}
where \( \mathcal{N}(y | \mu_i, \Sigma_i) \) is the multivariate normal distribution with mean vector \( \mu_i \) and covariance matrix \( \Sigma_i \). This capability allows MGMMs to accurately capture complex data structures where variables are interdependent, providing a more realistic representation of real-world data distributions \cite{mclachlan2004finite}. Unlike univariate models that assume independence, MGMMs are particularly effective in scenarios requiring flexible and scalable modeling of multidimensional data, such as in image processing \cite{Bueno2006}. By accommodating these correlations, MGMMs enhance clustering and classification tasks, enabling more meaningful groupings in several applications where multiple correlated features influence outcomes. Furthermore, MGMMs excel in accurate density estimation for multivariate data, which is crucial in fields like environmental science for modeling spatial distributions of pollutants or genetics to analyze complex gene expression profiles.

This work involves the use of multivariate data containing intrinsic correlations between the variables, making MGMMs one of the best options to accurately estimate the density of our data. By leveraging the ability of MGMMs to model these correlations through covariance matrices, we can achieve a more realistic and precise representation of the data distribution, which is crucial for our analysis. Furthermore, we can enhance our modeling capabilities by combining the approximation properties of neural networks with the flexibility of mixture models~\citep{Bishop1994}. This approach allows us to model complex density estimations without requiring any prior knowledge of their distributions.

\section{Mixture Density Networks}

Neural networks are computer models inspired by the structure of the human brain~\citep{Hornik1989}. They are made up of layers of connected neurons or nodes. Such layers are used to process input data, with each neuron applying an activation function and a weighted sum to produce an output. Through training, neural networks can discover intricate patterns and relationships in data.

However, in problems involving continuous variables where the same input values may produce different output values, neural networks tend to predict the mean of the target variable. This can be regarded as an approximation to the conditional average of the target variable given the input. This conditional average provides a very limited description of the statistical properties of the data and is often
inadequate for many applications. This is particularly true for non-unique inverse problems, where a conventional neural network with a least-squares approach might yield an inaccurate solution as the mean of multiple, possibly more accurate solutions.

In our case, averaging parameters such as \( R_{cn} \) and \( Z_{sn} \) tends to yield suboptimal results due to their complex interdependencies. Both variables exhibit multimodal distributions centered around symmetric values. Averaging these values tends to converge toward zero, which may lead to the generation of bad stellarator designs. Consequently, there is a need for a neural network to be capable of probabilistically selecting \( R_{cn} \) or \( Z_{sn} \) from their respective subdistributions, depending on the context. This leads to the use of a probabilistic model capable of representing multimodal distributions. Such a model would not average the distributions but instead sample from them, thereby preserving the distinct characteristics of each mode and enabling more accurate predictions and good stellarator designs.

To address these requirements, Mixture Density Networks (MDNs)~\citep{Bishop1994} present a compelling solution. MDNs are a class of neural networks designed to overcome the limitations of conventional neural networks in modeling complex, multi-modal data distributions. They combine the flexibility of neural networks with the robustness of mixture models, where the neural network estimates the parameters for the mixture model. MDNs allow a neural network to learn arbitrary conditional distributions as opposed to only learning the mean. This enables MDNs to provide a more comprehensive and accurate modeling approach for complex data distributions.

In MDNs, the probability density of the target data is represented as a linear combination of components, as in~\cref{eq:mixture_model}. Various choices for these components are possible, but for the purpose of this work, we focus on MGMMs, as in~\cref{eq:mgmm}, to approximate the conditional distribution of the target variables given the inputs,   because, as seen in \cref{section3c}, it effectively captures complex data structures where variables are interdependent, and excels in accurate density estimation for multivariate data, which is crucial for our case. 

\begin{figure}
    \centering
        \includegraphics[trim={4cm 3.5cm 6cm 2cm}, clip, width=.48\textwidth]{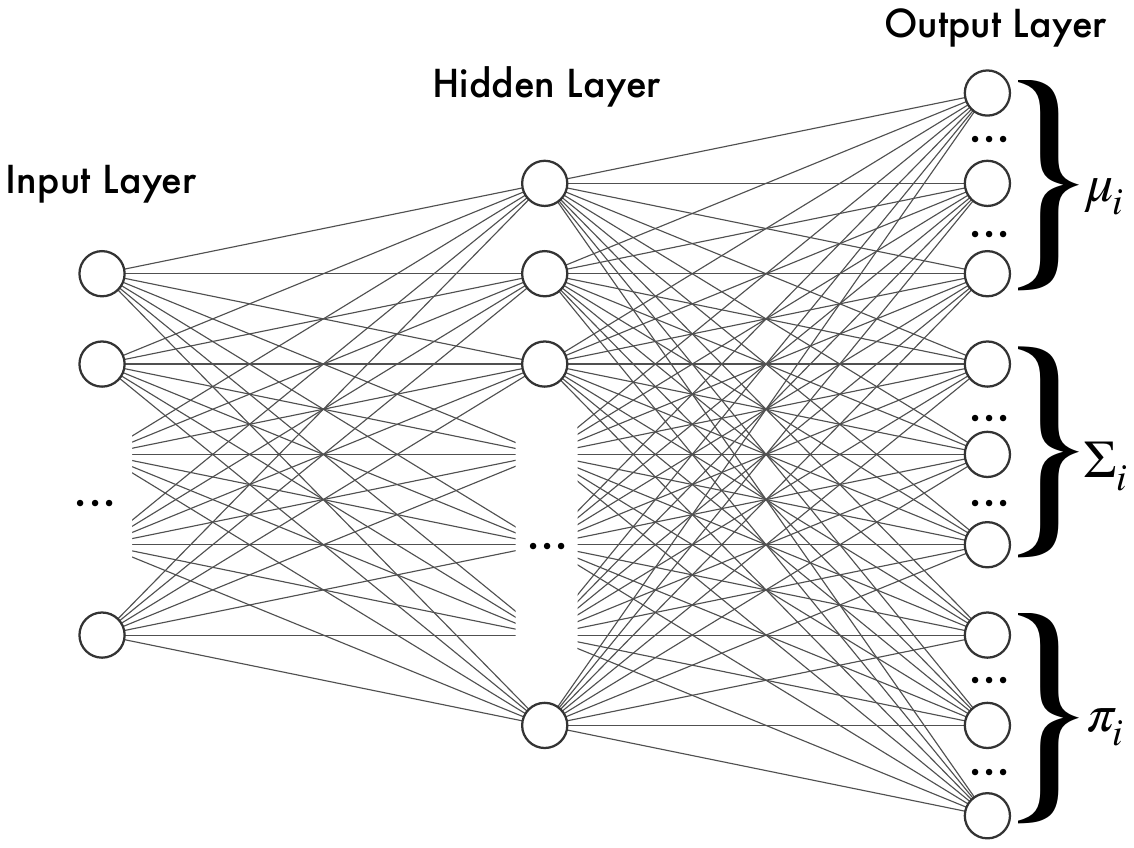}
        \includegraphics[trim={5cm 4.5cm 5cm 5cm}, clip, width=.48\textwidth]{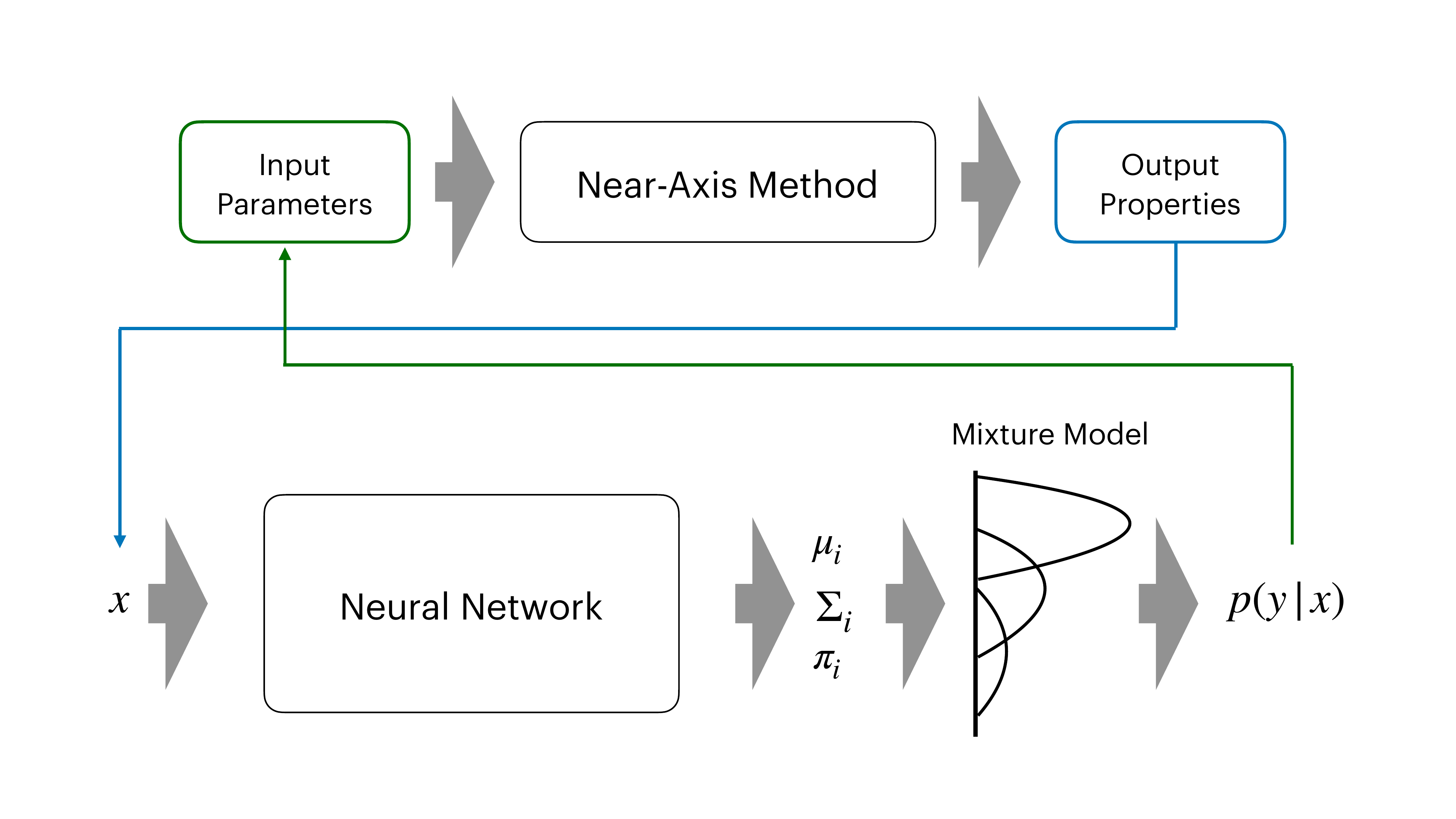}
    \caption{(\emph{left}) Sketch of the neural network architecture used in this work to estimate the parameters of a mixture model. (\emph{right}) Architecture of the Mixed Density Network as an inverse model for the near-axis method.}
    \label{fig:mdn_comparison}
\end{figure}

For any given values of the input \( x \), the MDN provides a systematic method for modeling an arbitrary conditional distribution \( p(y | x) \). The model parameters, namely the mixing coefficients \( \pi_i \), the mean vectors \( \mu_i \), and the covariance matrices \( \Sigma_i \), are modeled as continuous functions of \( x \). This is achieved by having \( \pi_i \), \( \mu_i \), \( \Sigma_i \) as the outputs of a conventional neural network, which takes \( x \) as its input. The combined structure of a feed-forward network and a mixture model is the essence of an MDN.  
The basic structure of the feedforward neural network responsible for modeling the parameters of the mixture as a continuous function of the input parameters is illustrated in~\cref{fig:mdn_comparison}. This architecture enables the network to dynamically adjust the mixture parameters based on the input data while capturing complex, nonlinear relationships in the data.

By choosing a mixture model with a large enough number of components, and a neural network with a large enough number of hidden units{~\citep{Uzair2020}}, the MDN can approximate any conditional density \( p(y | x) \) as closely as desired {~\citep{Lu2020}}. In this work, we use a mixture model with 62 components. This choice was empirically determined to provide an optimal balance between model complexity and performance. It was observed that increasing both the number of layers and the width of each layer, as well as incorporating more components, provided severe improvements in the model's performance. The architecture of the mixture density network used in this work is illustrated in~\cref{fig:mdn_comparison} and in~\cref{tab:tab_mdn_structure}, showcasing the detailed configuration and activation functions employed at various layers.

\begin{table} 
    \centering
    \begin{tabular}{ccc}
        \textbf{Layer} & \textbf{Size} & \textbf{Activation function} \\
        \hline
        Input & 10 & -- \\
        \hline
        Hidden 1 & 64 & tanh \\
        Hidden 2 & 128 & tanh \\
        Hidden 3 & 256 & tanh \\
        Hidden 4 & 512 & tanh \\
        Hidden 5 & 1024 & tanh \\
        Hidden 6 & 2048 & tanh \\
        \hline
        Output & 4092 & \begin{tabular}[c]{@{}c@{}}tanh for $\mu$ and $\Sigma_{ij,j>i}$, \\ ELU + 1 for $\Sigma_{ij,i=j}$, \\ softmax for $\pi$\end{tabular} \\
    \end{tabular}
    \caption{Layers of the Mixture Density Network used in this work.}
    \label{tab:tab_mdn_structure}
\end{table}

The neural network's input layer contains 10 neurons, corresponding to the 10 input parameters. This is followed by a series of hidden layers with progressively increasing sizes, namely 64, 128, 256, 512, 1024, and 2048. The output layer consists of 4092 nodes that are allocated as follows: 62 nodes represent the mixture weights of the 62 components; 620 nodes represent the mean vector of the 10 outputs for each of the 62 components; 3410 nodes represent the 55 parameters (the upper triangular part) of the \(10 \times 10\) covariance matrix for each of the 62 components.

The calculation for the number of parameters for each covariance matrix uses the formula \( D(D + 1)/2 \), where \( D \) is the dimension of the covariance matrix. With \( D = 10 \), each covariance matrix requires 55 parameters, resulting in a total of 3410 parameters for the 62 components. We employ the hyperbolic tangent, $\tanh$,  activation function to the hidden layers to prevent numerical issues that may arise from large values propagating through the network, which could lead to unstable computations and vanishing gradients, causing non-positive definite covariance matrices.

In the output layer, the neural network uses different activation functions tailored to the nature of each parameter type. The means of the Gaussian components are mapped to the range \(\interval[open]{-1}{1}\) using the $\tanh$ activation function, benefiting from the data normalization process (a standard scaler) which is applied to the input data, i.e., when normalized, we expect the data to become centered around zero, which agrees with the zero-centered nature of the $\tanh$ activation function and, at the same time, limits the potential for large numerical values propagating through the network. The mixture weights are computed using the softmax function~\citep{Bridle1990, Jacobs1991} to ensure that they sum to 1. The covariance matrices are computed with the diagonal elements using a modified ELU~\citep{Clevert2016} function (ELU + 1) function to ensure positivity, and the off-diagonal elements are constrained between \(\interval[open]{-1}{1}\) using the tanh activation function, for the same reasons presented before. With this established architecture, the next step involves training the MDN on a dataset of stellarator configurations to adjust the network weights, thereby enhancing the model’s predictive capabilities.

\section{Data Generation and Training}
\label[section]{sec:iterative_training}

To train the MDN, we generate a dataset of stellarators using the near-axis expansion method. The dataset is a collection of records containing the input parameters provided to the near-axis method and corresponding output properties generated from these inputs. These are listed in \cref{tab:input_parameters,tab:output_properties}.

To generate the dataset, we sample the input parameters from uniform distributions, with the ranges listed in~\cref{tab:generator_uniform} and find the output parameters listed in~\cref{tab:output_properties}.
{The range of parameters $R_{c2},R_{c3},Z_{c2}$ and $Z_{c3}$ follows the empirical observation in previous near-axis configurations and in the parameter scans done here that the Fourier coefficients generally decrease with increasing order. This allows us to restrict the database to feasible designs.}
By sampling the input parameters from uniform distributions, we find that most configurations consist of bad stellarators. In fact, by applying the set of criteria shown in~\cref{tab:good_stellarators}, it is seen that the percentage of good stellarators is extremely low, with only 1 in approximately 100,000 samples found to comply with all the desired criteria.
This illustrates how difficult it is to find good stellarators by random search, and is one of the main drivers of the use of an inverse model to find the input parameters from a set of desired properties.

\begin{table}
    \centering
    \begin{tabular}{lr}
        \textbf{Input Parameter} & \textbf{Range} \\
        \hline
        $R_{c1}$ &  $\interval{-1}{1}$ {($=\interval{-R_{c0}}{R_{c0}}$)} \\
        $R_{c2}$ & $\interval{-|R_{c1}|}{|R_{c1}|}$ \\
        $R_{c3}$ & $\interval{-|R_{c2}|}{|R_{c2}|}$ \\
        $Z_{s1}$ & $\interval{-1}{1}$  {($=\interval{-R_{c0}}{R_{c0}}$)} \\
        $Z_{s2}$ & $\interval{-|Z_{s1}|}{|Z_{s1}|}$ \\
        $Z_{s3}$ & $\interval{-|Z_{s2}|}{|Z_{s2}|}$ \\
        $|\bar{\eta}|$ & $\interval{0.01}{3.0}$ \\
        $|B_{2C}|$ & $\interval{0.01}{3.0}$ \\
        $n_{\text{fp}}$ & $\interval{0}{10}$ \\
        $p_2$ & $\interval{-4\times 10^6}{0.0}$ \\
    \end{tabular}
    \caption{Uniform distributions defining the input parameter ranges used for dataset generation. Each parameter is sampled within the interval shown in the second column.}
    \label{tab:generator_uniform}
\end{table}

Following the generation of the dataset, we begin by normalizing the dataset using a standard scaler to account for the different scales of the input and output parameters. The dataset was then split into training and validation sets with an 80\% and 20\% split, respectively.
Next, we initialize the weights of the neural network using the Glorot-Xavier initialization method~\citep{Glorot10}, which is effective for deep neural networks as it helps prevent vanishing or exploding gradients during training. Additionally, we employ the Adam optimizer~\citep{Adam17} with a learning rate of \(10^{-3}\) and a batch size of 10,000 samples.

The output properties are then sampled from the mixture model. We compute the negative log-likelihood of these samples, which serves as the loss function to be minimized during training. Since we use a mixture model composed of multiple Gaussian components, the loss function is given by
\begin{equation}
    \label{eq:loss}
    \text{Loss} = -\frac{1}{N} \sum_{j=1}^{N} \log \left( \sum_{i=1}^{K} \pi_i \mathcal{N}(y_j | \mu_i, \Sigma_i) \right),
\end{equation}
where \( N \) is the number of samples, i.e., the batch size, and \(y_j\) is an output vector.

\begin{figure}
    \centering
        \includegraphics[width=.48\textwidth]{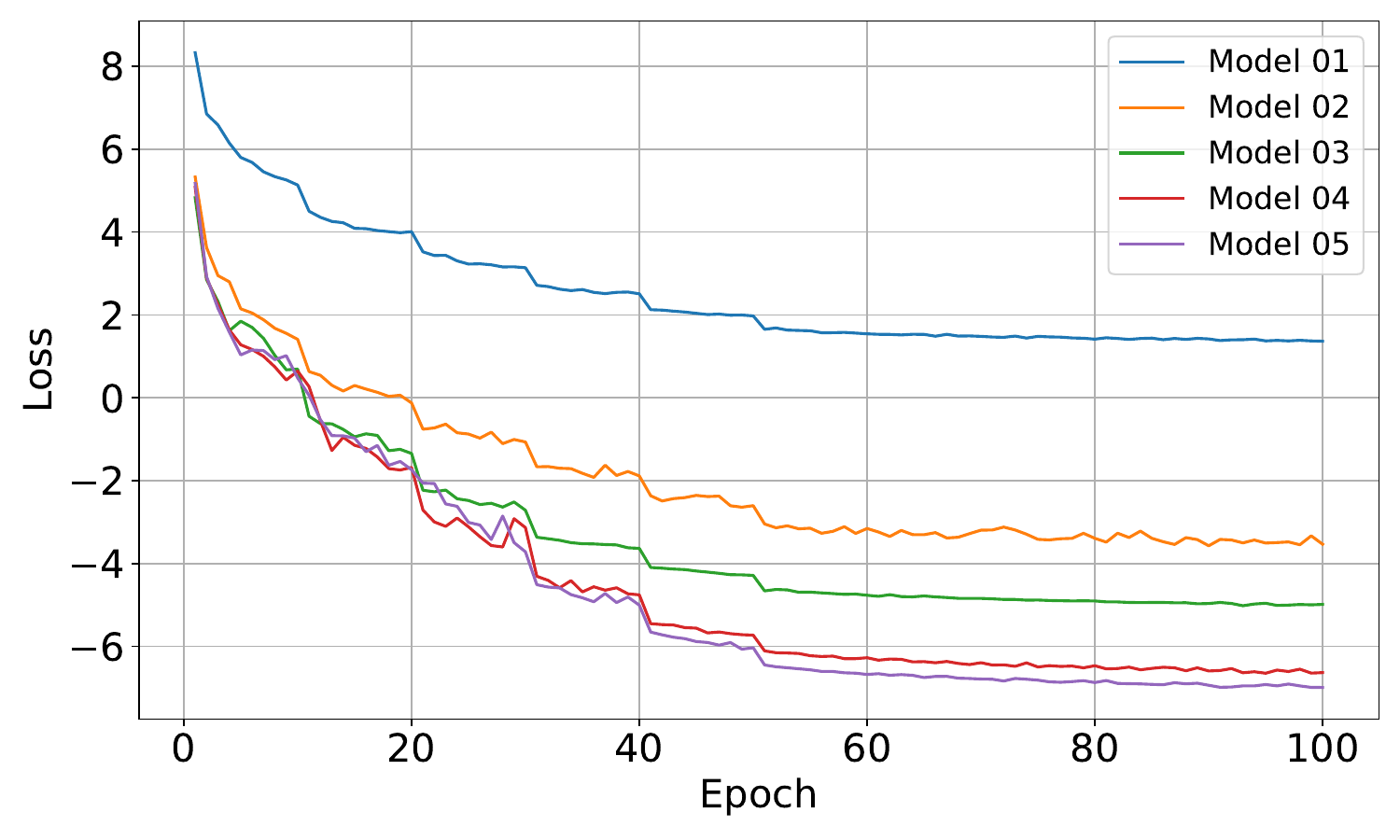}
        \includegraphics[width=.48\textwidth]{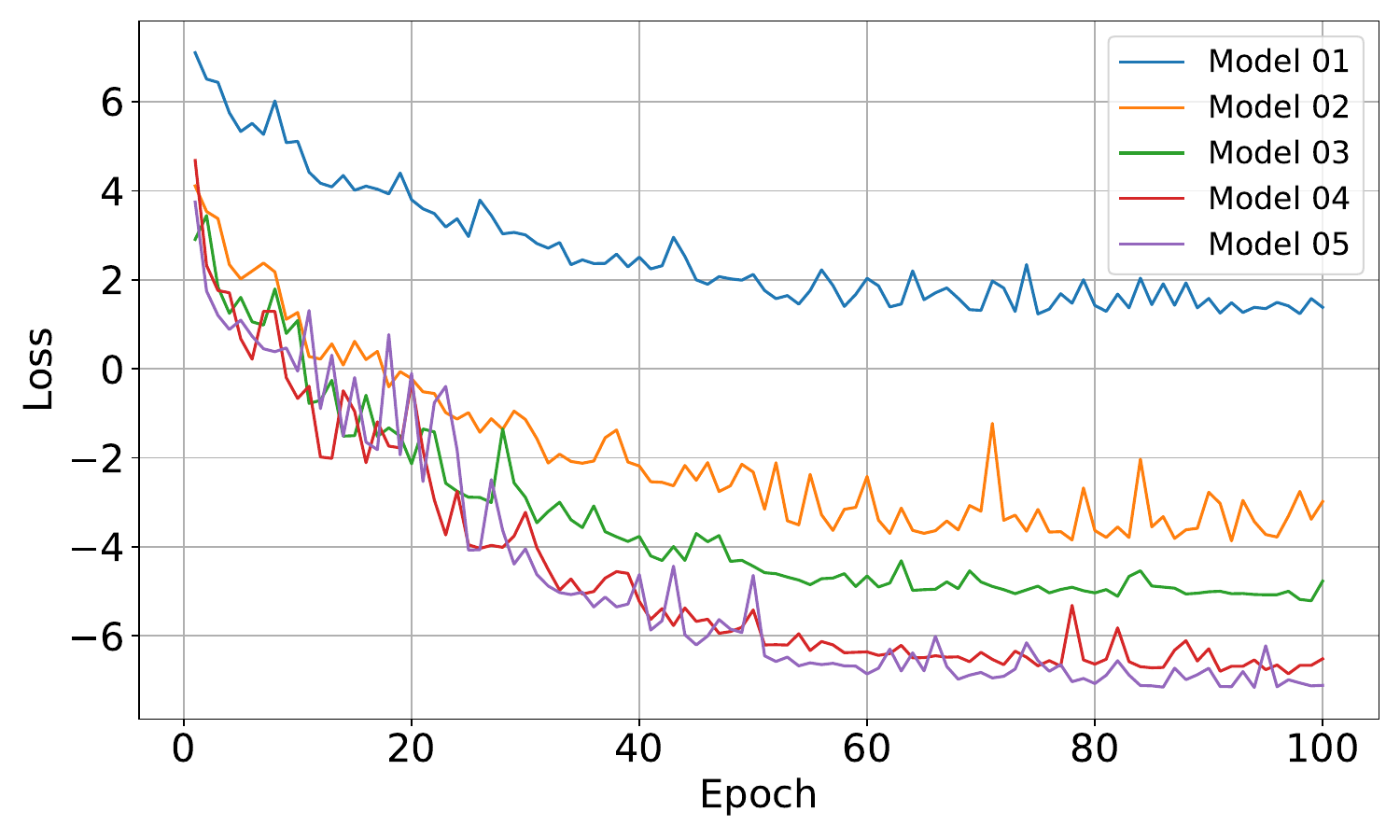}
    \caption{Loss (\emph{left}) and validation loss (\emph{right}) curves during training for the different models. The initial learning rate, $1 \times 10^{-3}$, was decreased with a scheduler in epochs 10, 20, 30, 40, 50 with a $\gamma = 0.5$.}
    \label{fig:loss_comparison}
\end{figure}

Despite using the Adam optimizer~\citep{Adam17}, the training process was more challenging than anticipated due to numerical instabilities, such as vanishing gradients, that caused the covariance matrices to become non-positive definite. To address this issue, {a multi-step learning rate scheduler was employed, which adjusted the learning rate at specific training epochs (10, 20, 30, 40, and 50) by a factor of 0.5. This schedule initially allowed the model to explore the parameter space with a higher learning rate, then gradually refined as training progressed. By reducing the learning rate in steps, the model avoided abrupt changes in parameter updates, leading to a more stable convergence}. The loss and validation curves can be seen in~\cref{fig:loss_comparison}. Notably, the curves indicate that as the learning rate decreases, the loss function values also decrease. This trend suggests that lower learning rates contribute to a more stable and gradual convergence, resulting in better model performance and lower loss.

However, as mentioned earlier, the percentage of good stellarators obtained by random sampling was very low. To address this issue, we adopted an iterative training approach, where the trained model was used to support the generation of a new dataset. This new dataset can be used to re-train the model, which in turn can be used to support the generation of a further dataset.

The uniform distributions in~\cref{tab:good_stellarators} have been used only once to generate the initial dataset. Once the model is trained, we use it to draw samples of input parameters of good stellarators to then provide to the near-axis method. At first, the model only had a small number of good stellarators (0.04\% after the first training). However, over the course of several training iterations, the percentage of good stellarators in the dataset keeps increasing. This is shown in Table~\ref{tab:all_data_good_stellarators_percentage} where, at the end of the fifth iteration, the percentage of good stellarators reaches approximately 20\%.
The resulting model is analyzed in the next section.

\begin{table} 
    \centering
    \begin{tabular}{lr}
        \textbf{Dataset} & \textbf{Good Stellarators (\%)} \\
        \hline
        Before training (uniform sampling) & 0.0018 \\
        After the first training iteration & 0.0406 \\
        After the second training iteration & 1.3788 \\
        After the third training iteration & 9.0024 \\
        After the fourth training iteration & 12.3903 \\
        After the fifth training iteration & 20.2670 \\
    \end{tabular}
    \caption{Percentage of good stellarators in each iteration dataset.}
    \label{tab:all_data_good_stellarators_percentage}
\end{table}

\begin{figure}
        \includegraphics[width=.48\textwidth]{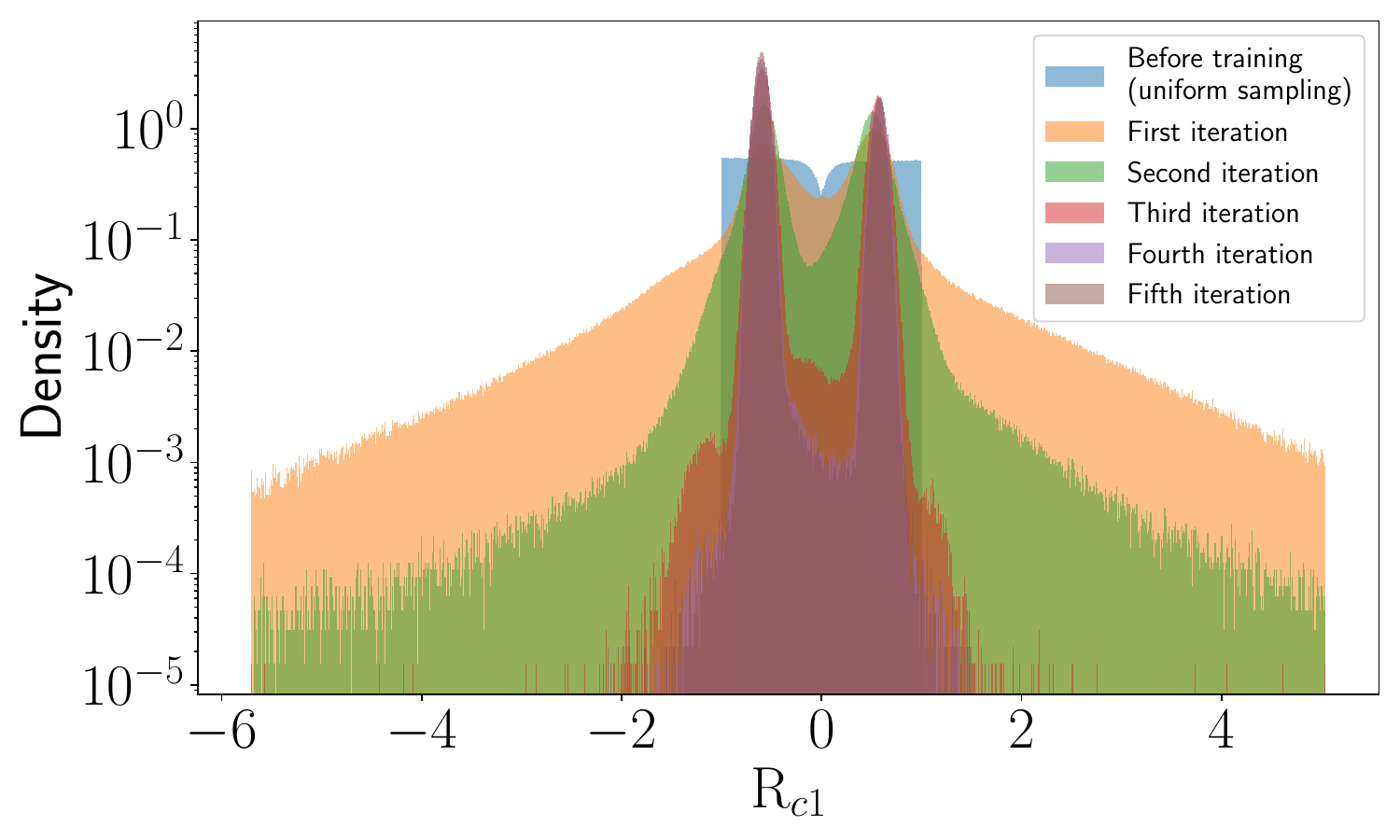}
        \includegraphics[width=.48\textwidth]{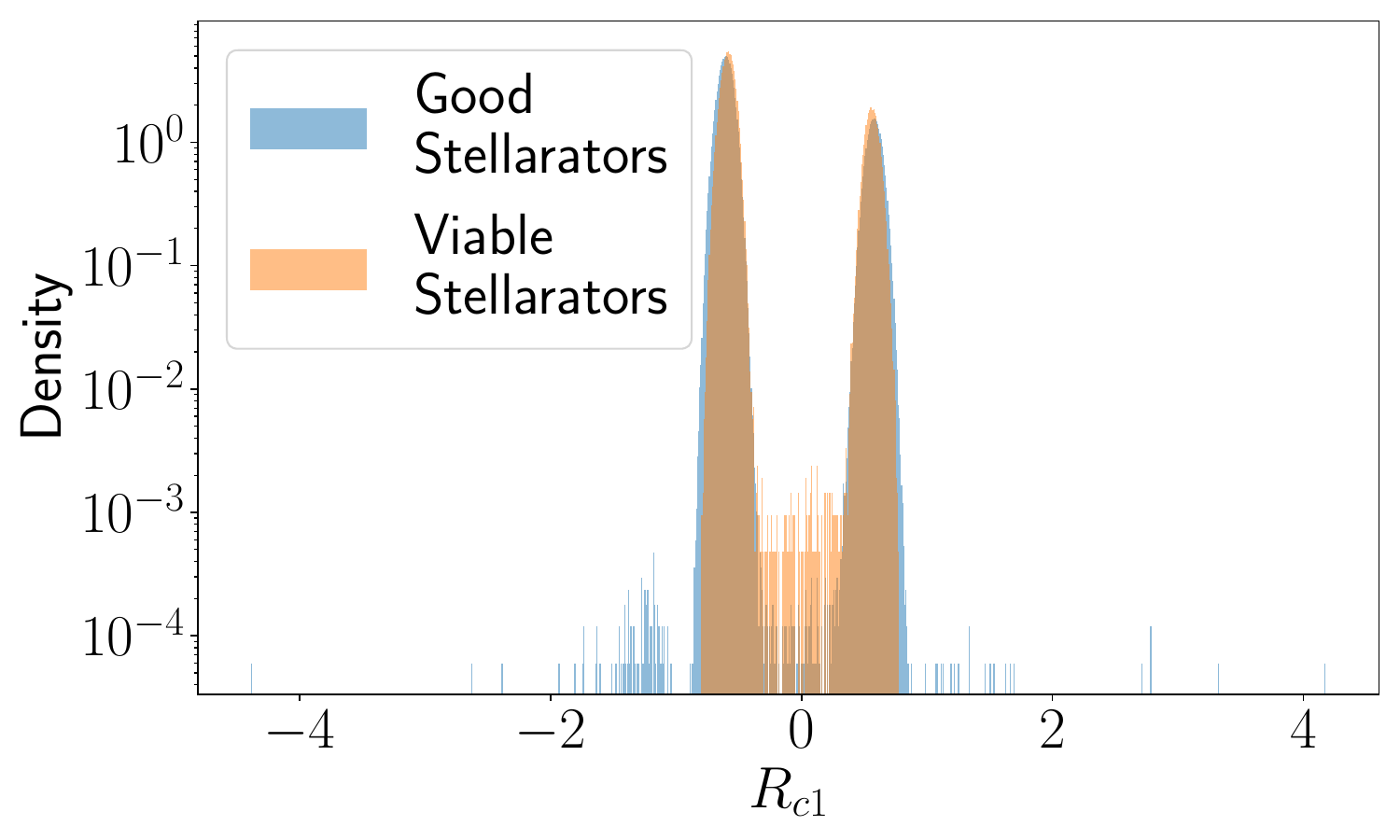}
    \caption{(\emph{left}) Distribution of the $R_{c1}$ variable during the iterative process. (\emph{right}) Distribution of the $R_{c1}$ variable for the good stellarators and the viable stellarators.}
    \label{fig:rc1_comparison}
\end{figure}

The evolution of the distribution of the $R_{c1}$ variable during the training of the model is shown in~\cref{fig:rc1_comparison}.
The initial uniform distribution used to create the dataset gradually transitions to a bimodal Gaussian-like distribution. This transformation aligns more closely with our objective of focusing on the region where good stellarators are found. This transition also simplified the training of the model, as Gaussian mixture models can more effectively approximate it compared to a uniform distribution, which would require more components with wider covariances.
Here, we find that the final distribution of the $R_{c1}$ variable has two peaks, one around -0.8 and another around 0.8, with a higher peak at -0.8.

\section{Model Performance}

We now show how the model can be used to predict the input parameters needed to obtain optimized stellarators with desired output properties.
First, the user provides the desired properties such as {the volume-averaged} plasma $\beta$ and rotational transform, and the model produces the design parameters that are likely to yield those properties such as magnetic axis and $\bar{\eta}$. Then, the user feeds the predicted design parameters to the near-axis expansion method, to generate the corresponding properties. Finally, the user verifies that the actual properties generated by the near-expansion method agree with the desired properties. {A randomly selected example from the dataset is presented in~\cref{tab:example}, while an example using the frontier conditions from~\cref{tab:good_stellarators} is shown in~\cref{tab:example_on_edge}.}

\begin{table} 
    \centering
    \begin{tabular}{lc|lc|c}                                                                     
        \multicolumn{2}{c|}{\textbf{Desired properties}} & \multicolumn{2}{|c|}{\textbf{Design parameters}} & \textbf{Actual properties} \\
        \multicolumn{2}{c|}{(input to MDN)} & \multicolumn{2}{|c|}{(output of MDN /} & (output of pyQSC) \\
        \multicolumn{2}{c|}{\textbf{}} & \multicolumn{2}{|c|}{input to pyQSC)} &  \\
        \hline
        axis length    & 12.23       & $R_{c1}$    & -0.492168  & 12.67    \\
        $\iota$            & -2.24       & $R_{c2}$    & 0.003776  & -2.07  \\
        max elongation & 6.56      & $R_{c3}$    & -0.000132 & 8.20   \\
        $\text{min }L_{\nabla B}$ & 0.41      & $Z_{s1}$    & -0.652899 & 0.44   \\
        min R0         & 0.48     & $Z_{s2}$    & 0.006861 & 0.51  \\
        $R_{\text{singularity}}$  & 0.14     & $Z_{s3}$    & -0.005334 & 0.11  \\
        $L_{\nabla\nabla B}$ & 0.25       & $n_{\text{fp}}$    & 3    & 0.28        \\
        $B_{20_{\text{variation}}}$  & 1.68      & $\bar\eta$ & -0.844595 & 3.94    \\
        $\beta$            & 0.005     & B2c    & 1.662730 & 0.003    \\
        $D_{\text{Merc}} \times r^2$ & 0.09      & p2     & -162627 & -0.15     \\
    \end{tabular}
    \caption{Sample results for given desired properties. A random stellarator configuration was selected from the test dataset, and its properties were used as input to the model to predict the design parameters. These predicted design parameters were then fed into the Near-Axis Method, which returned the actual properties. The resulting actual properties closely matched the desired ones.}
    \label{tab:example}
\end{table}

\begin{table} 
    \centering
    \begin{tabular}{lc|lc|c}                                                                     
        \multicolumn{2}{c|}{\textbf{Desired properties}} & \multicolumn{2}{|c|}{\textbf{Design parameters}} & \textbf{Actual properties} \\
        \multicolumn{2}{c|}{(input to MDN)} & \multicolumn{2}{|c|}{(output of MDN /} & (output of pyQSC) \\
        \multicolumn{2}{c|}{\textbf{}} & \multicolumn{2}{|c|}{input to pyQSC)} &  \\
        \hline
        axis length    & 0.00      & $R_{c1}$    & 0.614602  & 7.99    \\
        $\iota$            & 0.20       & $R_{c2}$    & -0.058358   & -0.79  \\
        max elongation & 10.0      & $R_{c3}$    & 0.020746 & 14.65   \\
        $\text{min }L_{\nabla B}$ & 0.10      & $Z_{s1}$    & 0.804627  & 0.21   \\
        min R0         & 0.30     & $Z_{s2}$    & 0.013770 & 0.31  \\
        $R_{\text{singularity}}$  & 0.05     & $Z_{s3}$    & 0.013673 & 0.03  \\
        $L_{\nabla\nabla B}$ & 0.10       & $n_{\text{fp}}$    & 1    & 0.10        \\
        $B_{20_{\text{variation}}}$  & 5.00      & $\bar\eta$ & 0.509180 & 5.51   \\
        $\beta$            & 0.001     & B2c    & -1.257980 & 0.00014    \\
        $D_{\text{Merc}} \times r^2$ & 0      & p2     & -154446 & 0.10     \\
    \end{tabular}
    \caption{Sample results for given desired properties that were the boundary conditions in \cref{tab:good_stellarators}. The properties of the given stellarator were used as input to the model to predict the design parameters. These predicted design parameters were then fed into the Near-Axis Method, which returned the actual properties. The resulting actual properties closely matched the desired ones.}
    \label{tab:example_on_edge}
\end{table}

However, while the model is able to yield configurations that satisfy the requirements listed in \cref{tab:good_stellarators}, it is not guaranteed that all configurations have a set of nested, non-intersecting flux surfaces up to the parameter $r_{\text{singularity}}$.
This is because $r_{\text{singularity}}$ is only a proxy for the minimum aspect ratio of the device.
Only by computing the surface in Cartesian coordinates, as opposed to the near-axis Boozer coordinates used throughout this work, can we verify the existence of such a surface.
Such an evaluation is crucial to use such configurations in practice.
We then take all the good stellarators and generate a surface at a radial distance of $r=0.1 R_{c0}$.
Here, the existence of such a surface is defined as the existence of a numerical solution of the mapping from the toroidal Boozer coordinate $\varphi$ on-axis to a cylindrical angle $\phi$ off-axis with tolerance at or below $10^{-15}$ after a maximum of 1000 iterations.
We will refer to the good stellarators that meet this additional criterion as \emph{viable} stellarators.

Next, keeping the standard normalization on the dataset, we employed the Huber Loss and the Mean Absolute Error (MAE) as evaluation metrics to compare the predicted output properties from the model against the output properties from the near-axis model on 10,000 samples. Both are metrics used in regression tasks to quantify the difference between predicted values and actual observations. Huber Loss combines the advantages of MAE for robustness to outliers and Mean Squared Error (MSE) for sensitivity to small errors. The results for {bad, good and} viable stellarators are presented in~\cref{tab:results_extra_good}.

\begin{table} 
    \centering
    \begin{tabular}{ccccccc}
        \multirow{3}{*}{\textbf{Variable}} & \multicolumn{2}{c}{\textbf{Viable}} & \multicolumn{2}{c}{\textbf{Good}} & \multicolumn{2}{c}{\textbf{Bad}} \\

         & \multicolumn{2}{c}{\textbf{Metric}} & \multicolumn{2}{c}{\textbf{Metric}} & \multicolumn{2}{c}{\textbf{Metric}} \\ 
                                           & \textbf{Huber Loss} & \textbf{MSE} & \textbf{Huber Loss} & \textbf{MSE} & \textbf{Huber Loss} & \textbf{MSE} \\
        \hline
        axis length                         & 0.031 & 0.0618 & 0.0342 & 0.083 & 1.33 & 10.2 \\ 
        $\iota$                             & 0.0267 & 0.0534  & 0.0233 & 0.0495 & 0.909 & 4.2 \\ 
        max elongation                      & 0.000456 & 0.0113  & 0.000326 & 0.068 & 0.138 & 17.8 \\ 
        $\text{min }L_{\nabla B}$           & 0.266 & 0.735  & 0.227 & 0.578 & 0.869 & 3.82 \\ 
        $\text{min }R_0$                    & 0.00617 & 0.0531  & 0.00665 & 0.274 & 2.72 & 32.6 \\ 
        $r_{\text{singularity}}$            & 0.632 & 1.72  & 0.907 & 3.11 & 0.0098 & 0.0292 \\ 
        $L_{\nabla\nabla B}$                & 0.432 & 1.12  & 0.415 & 1.08 & 0.149 & 0.366 \\ 
        $B_{20_{\text{variation}}}$         & 0.000604 & 0.0046  & 8.99$\times 10^{-5}$ & 0.000218 & 3.9 & 37.4 \\ 
        $\beta$                             & 0.321 & 1.1  & 0.658 & 3.12 & 0.00454 & 0.0165 \\ 
        $D_{\text{Merc}}\times r^2$         & 3.4 $\times 10^{-11}$ & 5.94 $\times 10^{-5}$  & 1.65$\times 10^{-10}$ & 9.94$\times 10^{-12}$ & 83.7 & 124.3 \\ 
        \hline
        Average                             & 0.172 & 0.486  & 0.227 & 0.837 & 9.37 & 23.073 \\ 
        \hline
    \end{tabular}
    \caption{Model accuracy on {bad, good and} viable stellarators}
    \label{tab:results_extra_good}
\end{table}

As illustrated in~\cref{tab:results_extra_good} {for viable stellarators}, the model accuracy was found to be satisfactory. For the variables \textit{axis length}, $\iota$, \textit{max elongation}, $B_{20_{\text{variation}}}$, and $D_{Merc} \times r^2$, the model showed a good performance, evidenced by a low Huber and MSE losses, 0.172 and 0.486 respectively, with the MSE being higher than the Huber Loss, as expected.
Regarding the variables $\text{min }L_{\nabla B}$, $\text{min }R_0$, and $L_{\nabla\nabla B}$, the model displayed moderate accuracy under the Huber Loss metric. However, the MSE was higher, indicating that the model underperforms in these variables.
The variables $\beta$ and $r_{\text{singularity}}$ exhibited the poorest accuracy, with both metrics indicating suboptimal results. {A possible explanation for this outcome might be due to trade-offs in variable correlations, i.e., maximizing performance for some variables may require sacrificing accuracy in others.}
 
Beyond the model performance, understanding the relationships between variables is crucial for interpreting the behavior of output properties and their interdependencies. This knowledge significantly influences how the model should be used to predict input parameters. When output properties are strongly correlated, the model must carefully balance these correlations to achieve the desired outputs. Additionally, being aware of the distribution of variables is essential to ensure the model operates within familiar data spaces; otherwise, it may perform poorly. Therefore, analyzing the distributions of the variables and their correlations is vital.

\begin{figure}
        \includegraphics[width=.48\textwidth]{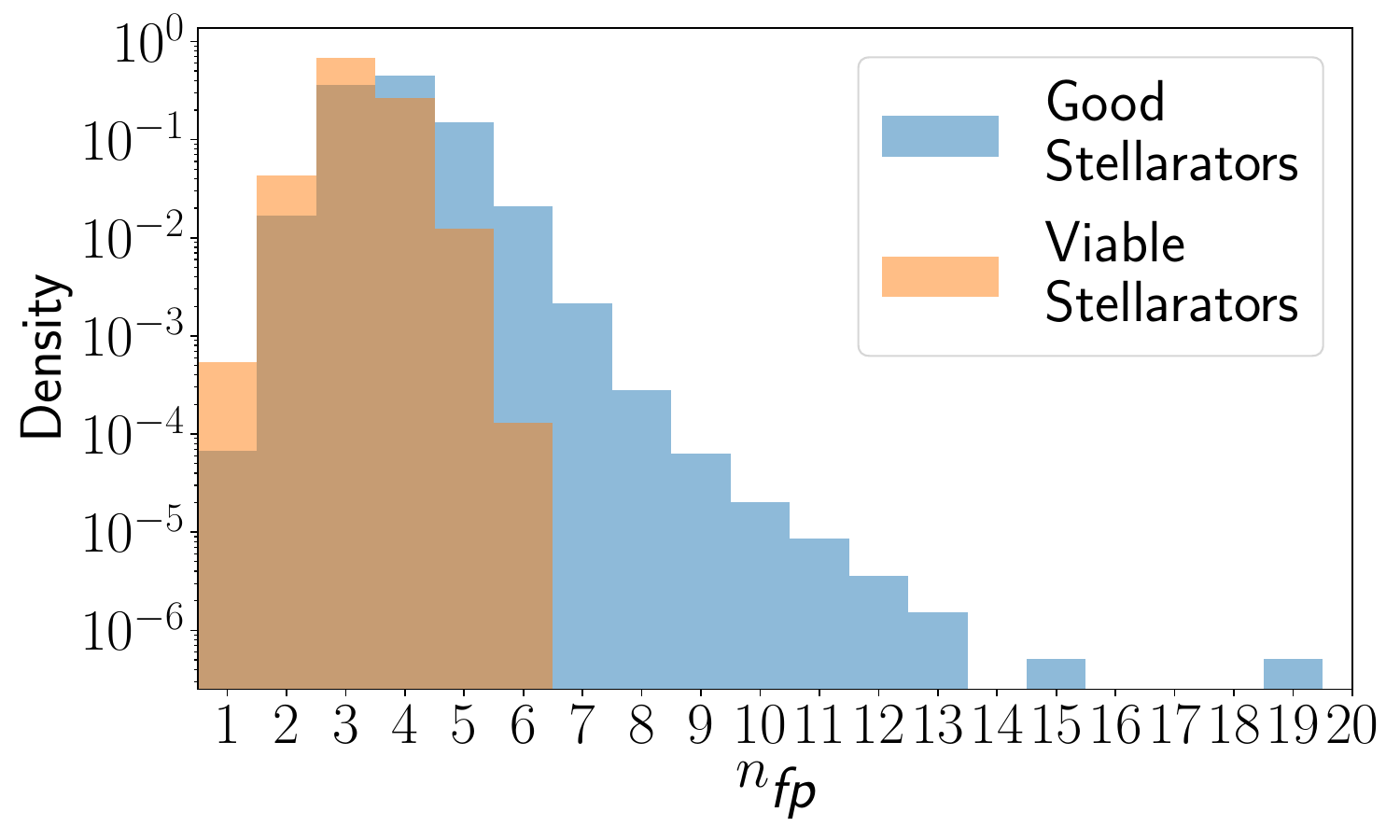}
        \includegraphics[width=.48\textwidth]{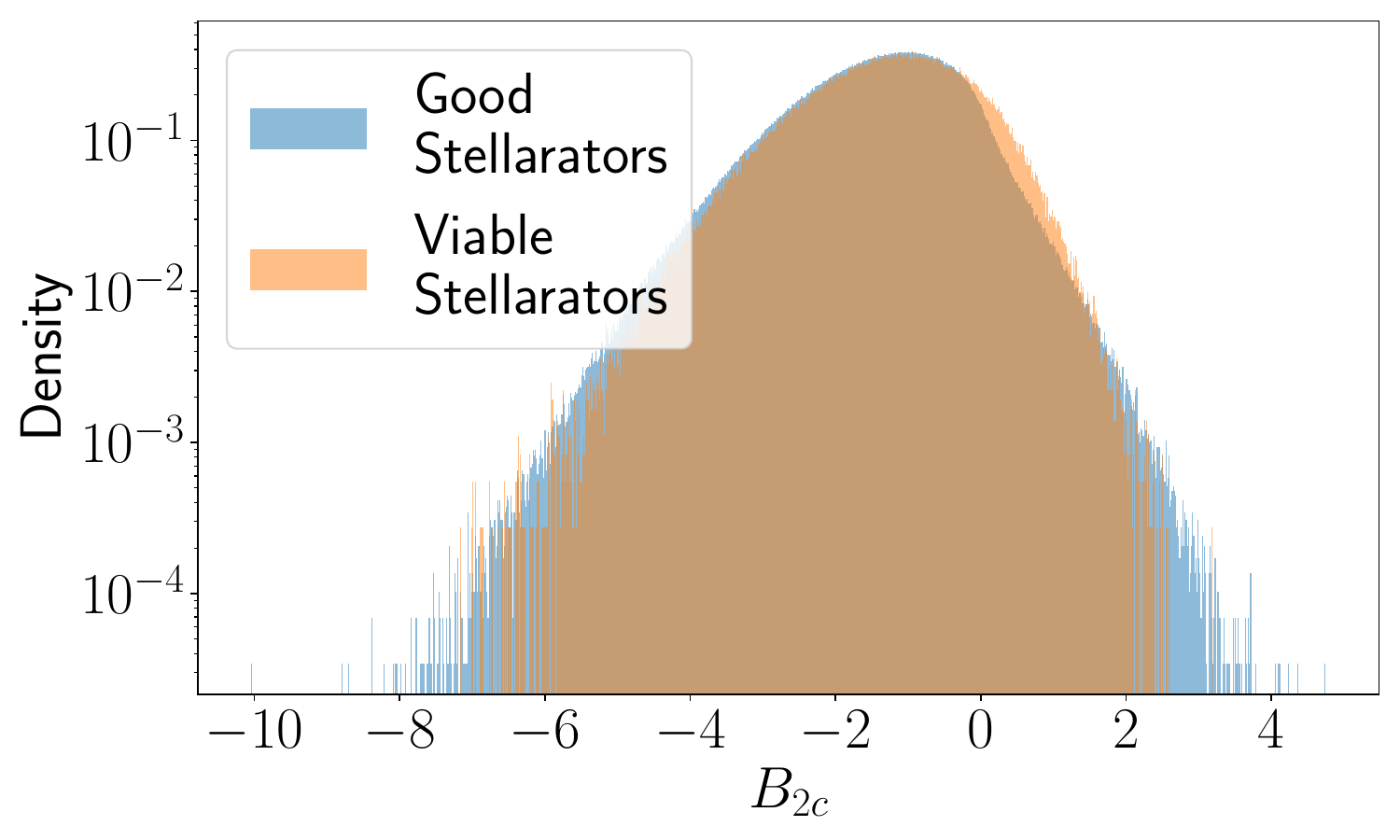}
    \caption{Distribution of number of field periods, $n_{\textit{fp}}$, (\emph{left}) and $B_{2c}$ variable (\emph{right}) for good and viable stellarators.}
    \label{fig:stellarator_distributions}
\end{figure}

Henceforth, the iterative training process described in~\cref{sec:iterative_training} was monitored to check if the distributions of both input and output variables were being restricted to a narrower space, which was to be expected since we wanted to restrict the dataset to the space of good stellarators. We evaluated the distribution of variables for the dataset containing all the good stellarators and all the viable stellarators. We show in~\cref{fig:rc1_comparison,fig:stellarator_distributions} the ones that provide a better understanding of the dataset and that are more relevant.

The distribution of the $n_{\text{fp}}$ variable for both the good stellarators and the viable stellarators is depicted in~\cref{fig:stellarator_distributions}. The data shows that good stellarators tend to cluster around $n_{\text{fp}}=4$, although there is a notable variation with several other $n_{\text{fp}}$ values present. An aspect of these results is that the model, despite being trained on a dataset where the $n_{\text{fp}}$ ranged from 1 to 10, successfully predicted $n_{\text{fp}}$ values for good stellarators that exceeded this range. As illustrated in~\cref{fig:stellarator_distributions}, there are configurations with $n_{\text{fp}}$ values extending up to 19.

For the viable stellarators, the distribution of the number of field periods, $n_{\text{fp}}$, is more narrowly centered around the value of 3, and none of the configurations exhibit $n_{\text{fp}}$ values above 6. This suggests a more constrained and specific range for $n_{\text{fp}}$ in the viable stellarator subset, indicating that these configurations are more consistent in this regard.
{The fact that an optimized stellarator with a higher number of field periods is hard to find, as it was also observed in \cite{Landreman_2022}, may be related to the fact that such $n_{\text{fp}}$ usually require a significant excursion of the axis and an associated larger axis length.
Furthermore, the recent study by \cite{Kappel2024} has shown a correlation between the number of field periods $n_p$
and $L_{\nabla B}$, indicating that small values of $n_p$ may lead to more optimized configurations.}

We also examine the $B_{2c}$ parameter. The distribution of this variable for both good stellarators and viable stellarators is illustrated in~\cref{fig:stellarator_distributions}. The data reveals that $B_{2c}$ exhibits a noticeable shift towards negative values. This indicates a distinct characteristic in the $B_{2c}$ distribution for good stellarators compared to the overall dataset.

\begin{figure} 
    \centering
    \includegraphics[width=0.7\textwidth]{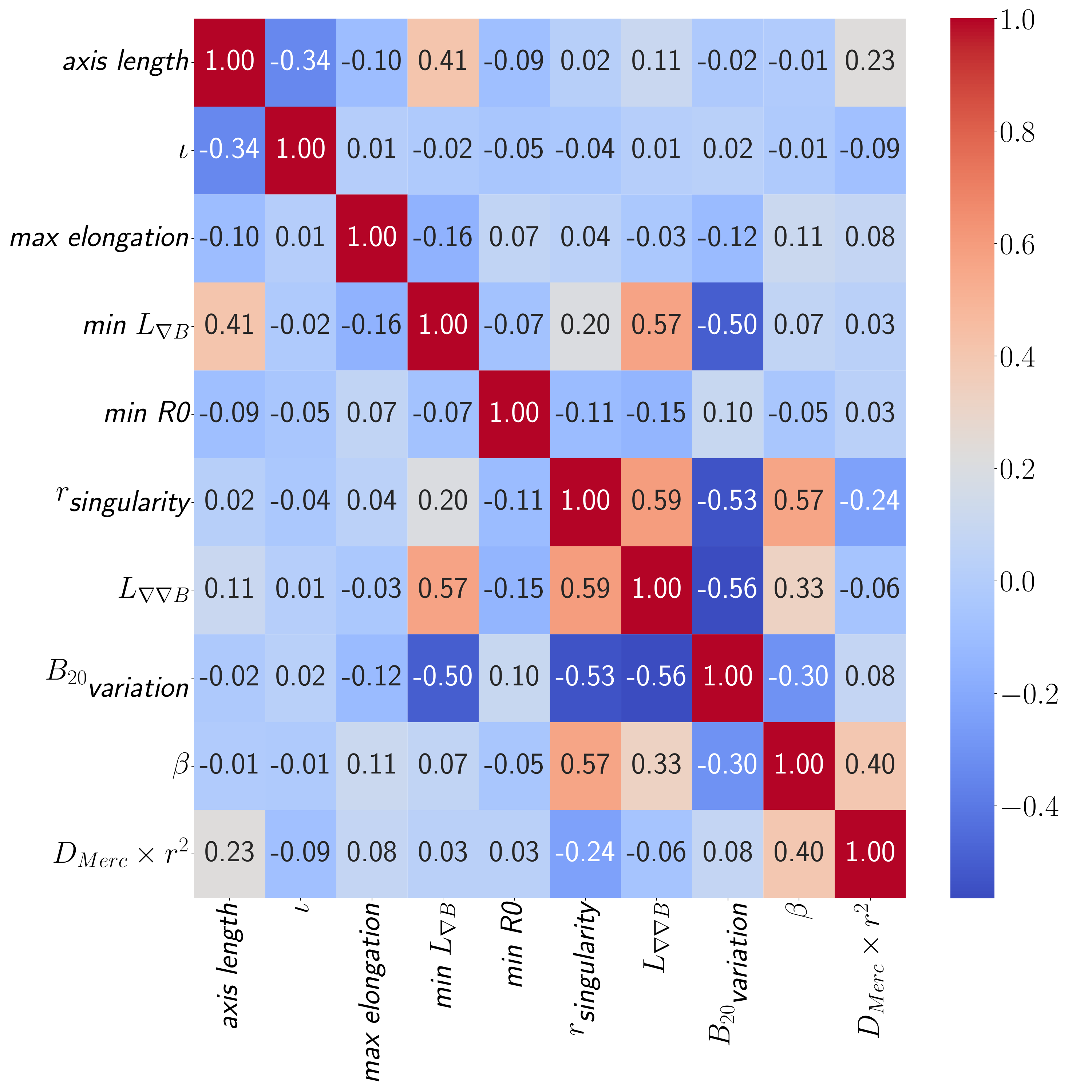}
    \caption{Correlation matrix for the output properties of good stellarators using Spearman Coefficient. The values range from -1 to 1, where negative values indicate negative correlations and positive values indicate positive correlations. The absolute values represent the correlation strength: values from 0 to 0.3 indicate a weak correlation, from 0.4 to 0.6 indicate a moderate correlation, and from 0.7 to 1 indicate a strong correlation.}
    \label{fig:correlation_matrix_good}
\end{figure}

The observed shifts in the distributions of variables for the good stellarators and the viable stellarators, whether towards negative or positive values, suggest that maximizing or minimizing certain variables can influence others in similar or opposing ways. This prompts us to evaluate the correlations between variables. While correlation does not imply causation, it provides valuable insights into the relationships between variables. We show in \cref{fig:correlation_matrix_good} the correlation matrix for the output properties of good stellarators, which is similar to viable stellarators. This matrix reveals a strong positive correlation between $r_{\text{singularity}}$ and $L_{\nabla \nabla B}$. This indicates that as the axis length increases, the maximum elongation also increases. Conversely, the $\text{min }L_{\nabla B}$ and $B_{20_{\text{variation}}}$ display a strong negative correlation, meaning that an increase in the minimum $\text{min }L_{\nabla B}$ results in a decrease in the minimum $B_{20_{\text{variation}}}$.
These relationships significantly impact model performance, as the model must balance them to achieve the desired properties.
As an example, if a user requests a stellarator with a high $\text{min }L_{\nabla B}$ and a low $B_{20_{\text{variation}}}$, the model must navigate the positive correlation between these properties. Since they are not independent, the model must find a compromise to generate appropriate input parameters that align with the desired output properties.

\section{Conclusions}

This work introduces an MDN designed to tackle the inverse stellarator optimization problem using the near-axis method. The model was trained on a dataset of near-axis configurations generated through the near-axis expansion method. However, the dataset initially contained a very low percentage of desirable stellarators, specifically only 0.001\%. To address this limitation, an iterative data augmentation technique was employed. This iterative approach successfully enhanced the representation of high-quality stellarators within the dataset, thereby improving the model's capability to predict parameters crucial for optimal stellarator designs.

Despite achieving good performance in predicting some variables, the model faced challenges with variables derived from the second-order near-axis expansion method, as assessed using Huber Loss and Mean Absolute Error (MAE) metrics. Nevertheless, overall, the MDN proved effective as a tool for predicting desired properties of stellarators.
{Our model can also return the covariance matrix to compute uncertainties associated with each prediction and obtain statistical insight.}

Moreover, the creation of a large database of high-quality stellarators facilitated detailed analyses of variable distributions and correlations. These analyses revealed that optimal stellarators tend to cluster within specific ranges of variable space, such as an $n_{\text{fp}}$ value around 3 or 4, and a preference for negative values in $B_{2c}$. The correlation matrix further highlighted strong interdependencies among variables, crucial for accurately predicting input parameters to achieve desired output properties.

As a future work, {an ablation study would be crucial to simplify the model, as the increasing complexity of the hidden layer geometry may not be optimal. Adding to this,} we intend to integrate the near-axis expansion method directly into the neural network training process, potentially as a differentiable layer. This advancement could leverage techniques like neural network approximations or automatic differentiation tools such as JAX~\citep{jax2018github}. Such enhancements would support the adoption of variational autoencoders, graph neural networks, and transformers. Such models could also be extended for future optimizations and designs, integrating them with an ideal MHD model rather than relying solely on a near-axis method. Additionally, a model could be developed to map between the near-axis method and an ideal MHD model. This approach would enable leveraging machine learning models for solving the inverse problem using a near-axis method and subsequently mapping the results to a full ideal MHD optimization.

\section*{Acknowledgments}
We would like to thank Raheem Hashmani and Misha Padidar for their insightful discussions throughout this work. 
%
R. Jorge would like to acknowledge the support of EUROfusion through an Enabling Research Grant, and the support of FCT -- \textit{Funda\c{c}\~{a}o para a Ci\^{e}ncia e Tecnologia, I.P.} through project reference \href{https://doi.org/10.54499/2021.02213.CEECIND/CP1651/CT0004}{2021.02213.CEECIND/CP1651/CT0004}.
This material is based upon work supported by the National Science Foundation under Grant No. 2409066.
This work has been carried out within the framework of the EUROfusion Consortium, funded by the European Union via the Euratom Research and Training Programme (Grant Agreement No 101052200 — EUROfusion). Views and opinions expressed are however those of the author(s) only and do not necessarily reflect those of the European Union or the European Commission. Neither the European Union nor the European Commission can be held responsible for them.
This work used Jetstream2 at Indiana University through allocation PHY240054 from the Advanced Cyberinfrastructure Coordination Ecosystem: Services \& Support (ACCESS) program, which is supported by National Science Foundation grants \#213859, \#2138286, \#2138307, \#2137603 and \#2138296.
This research used resources of the National Energy Research
Scientific Computing Center, a DOE Office of Science User Facility
supported by the Office of Science of the U.S. Department of Energy
under Contract No. DE-AC02-05CH11231 using NERSC award
NERSC DDR-ERCAP0030134.
This research used resources of the Oak Ridge Leadership Computing Facility at the Oak Ridge National Laboratory, which is supported by the Office of Science of the U.S. Department of Energy under Contract No. DE-AC05-00OR22725.
IPFN activities were supported by FCT - Fundação para a Ciência e Tecnologia, I.P. by project reference UIDB/50010/2020 and DOI identifier 10.54499/UIDB/50010/2020, by project reference UIDP/50010/2020 and DOI identifier 10.54499/UIDP/50010/2020 and by project reference LA/P/0061/2020 and DOI identifier 10.54499/LA/P/0061/2020.

\section*{Supplementary Data}

Supplementary material is available at \href{https://zenodo.org/records/13623959}{https://zenodo.org/records/13623959}.

\section*{Declaration of Interest}

The authors report no conflict of interest. 

\section*{Data Availability Statement}

The data that support the findings of this study are openly available in MLStellaratorDesign at \href{https://github.com/pedrocurvo/MLStellaratorDesign}{https://github.com/pedrocurvo/MLStellaratorDesign}.

\bibliographystyle{jpp}

\bibliography{references}

\begin{thebibliography}{34}
\expandafter\ifx\csname natexlab\endcsname\relax\def\natexlab#1{#1}\fi
\def\au#1{#1} \def\ed#1{#1} \def\yr#1{#1}\def\at#1{#1}\def\jt#1{\textit{#1}} \def\bt#1{#1}\def\bvol#1{\textbf{#1}} \def\vol#1{#1} \def\pg#1{#1} \def\publ#1{#1}\def\arxiv#1{#1}\def\org#1{#1}\def\st#1{\textit{#1}}

\bibitem[Bader {\em et~al.\/}(2019)Bader, Drevlak, Anderson, Faber, Hegna, Likin, Schmitt \& Talmadge]{Bader2019}
{\sc \au{Bader, A.}, \au{Drevlak, M.}, \au{Anderson, D.~T.}, \au{Faber, B.~J.}, \au{Hegna, C.~C.}, \au{Likin, K.~M.}, \au{Schmitt, J.~C.} \& \au{Talmadge, J.~N.}} \yr{2019}  \at{{Stellarator equilibria with reactor relevant energetic particle losses}}.  \jt{Journal of Plasma Physics}  \bvol{85}~(5),  \pg{905850508}.

\bibitem[Bishop(1994)]{Bishop1994}
{\sc \au{Bishop, C.}} \yr{1994}  \bt{Mixture density networks}. {\em Tech. Rep.\/} NCRG/94/004.  \org{Aston University}.

\bibitem[Boozer(2020)]{Boozer_2020}
{\sc \au{Boozer, A.}} \yr{2020}  \at{Why carbon dioxide makes stellarators so important}.  \jt{Nuclear Fusion}  \bvol{60}~(6),  \pg{065001}.

\bibitem[Boozer(1981)]{Boozer1981}
{\sc \au{Boozer, A.~H.}} \yr{1981}  \at{{Plasma equilibrium with rational magnetic surfaces}}.  \jt{Physics of Fluids}  \bvol{24}~(11),  \pg{1999}.

\bibitem[Bradbury {\em et~al.\/}(2018)Bradbury, Frostig, Hawkins, Johnson, Leary, Maclaurin, Necula, Paszke, VanderPlas, Wanderman-Milne \& Zhang]{jax2018github}
{\sc \au{Bradbury, J.}, \au{Frostig, R.}, \au{Hawkins, P.}, \au{Johnson, M.~J.}, \au{Leary, C.}, \au{Maclaurin, D.}, \au{Necula, G.}, \au{Paszke, A.}, \au{VanderPlas, J.}, \au{Wanderman-Milne, S.} \& \au{Zhang, Q.}} \yr{2018} {JAX: composable transformations of Python+NumPy programs}.

\bibitem[Bridle(1990)]{Bridle1990}
{\sc \au{Bridle, J.}} \yr{1990} Probabilistic interpretation of feedforward classification network outputs, with relationships to statistical pattern recognition.  \bt{In {\em Neurocomputing\/} (ed. \ed{F.F. Soulié \& J.~Hérault})},  \st{NATO ASI Series},  \vol{vol.~68}.

\bibitem[Bueno \& Kragic(2006)]{Bueno2006}
{\sc \au{Bueno, J.~I.} \& \au{Kragic, D.}} \yr{2006}  \at{Integration of tracking and adaptive gaussian mixture models for posture recognition}.  \jt{ROMAN 2006 - The 15th IEEE International Symposium on Robot and Human Interactive Communication}  \pg{pp. 623--628}.

\bibitem[Clevert {\em et~al.\/}(2016)Clevert, Unterthiner \& Hochreiter]{Clevert2016}
{\sc \au{Clevert, D.}, \au{Unterthiner, T.} \& \au{Hochreiter, S.}} \yr{2016}  \at{Fast and accurate deep network learning by exponential linear units ({ELUs})}.  \jt{arXiv 1511.07289} .

\bibitem[Cover \& Thomas(2012)]{Cover2012}
{\sc \au{Cover, T.~M.} \& \au{Thomas, J.A.}} \yr{2012} {\em Elements of Information Theory\/}.  \publ{Wiley}.

\bibitem[Garren \& Boozer(1991{\natexlab{{\em a\/}}})]{Garren1991a}
{\sc \au{Garren, D.~A.} \& \au{Boozer, A.~H.}} \yr{1991{\natexlab{{\em a\/}}}}  \at{{Existence of quasihelically symmetric stellarators}}.  \jt{Physics of Fluids B}  \bvol{3}~(10),  \pg{2822}.

\bibitem[Garren \& Boozer(1991{\natexlab{{\em b\/}}})]{Garren1991}
{\sc \au{Garren, D.~A.} \& \au{Boozer, A.~H.}} \yr{1991{\natexlab{{\em b\/}}}}  \at{{Magnetic field strength of toroidal plasma equilibria}}.  \jt{Physics of Fluids B}  \bvol{3}~(10),  \pg{2805}.

\bibitem[Glorot \& Bengio(2010)]{Glorot10}
{\sc \au{Glorot, X.} \& \au{Bengio, Y.}} \yr{2010} Understanding the difficulty of training deep feedforward neural networks.  \bt{In {\em Proceedings of the Thirteenth International Conference on Artificial Intelligence and Statistics\/}},  \st{PMLR},  \vol{vol.~9},  \pg{pp. 249--256}.

\bibitem[Goodfellow {\em et~al.\/}(2016)Goodfellow, Bengio \& Courville]{Goodfellow2016}
{\sc \au{Goodfellow, I.}, \au{Bengio, Y.} \& \au{Courville, A.}} \yr{2016} {\em Deep Learning\/}.  \publ{MIT Press}.

\bibitem[Helander(2014)]{Helander2014}
{\sc \au{Helander, P.}} \yr{2014}  \at{{Theory of plasma confinement in non-axisymmetric magnetic fields}}.  \jt{Reports on Progress in Physics}  \bvol{77}~(8),  \pg{087001}.

\bibitem[Hornik {\em et~al.\/}(1989)Hornik, Stinchcombe \& H.]{Hornik1989}
{\sc \au{Hornik, K.}, \au{Stinchcombe, M.} \& \au{H., White.}} \yr{1989}  \at{Multilayer feedforward networks are universal approximators}.  \jt{Neural Networks}  \bvol{2}~(5),  \pg{359--366}.

\bibitem[Jacobs {\em et~al.\/}(1991)Jacobs, Jordan, Nowlan \& Hinton]{Jacobs1991}
{\sc \au{Jacobs, R.}, \au{Jordan, M.}, \au{Nowlan, S.} \& \au{Hinton, G.}} \yr{1991}  \at{Adaptive mixtures of local experts}.  \jt{Neural Computation}  \bvol{3}~(1),  \pg{79--87}.

\bibitem[Jorge {\em et~al.\/}(2020)Jorge, Sengupta \& Landreman]{Jorge2020}
{\sc \au{Jorge, R.}, \au{Sengupta, W.} \& \au{Landreman, M.}} \yr{2020}  \at{{Near-axis expansion of stellarator equilibrium at arbitrary order in the distance to the axis}}.  \jt{Journal of Plasma Physics}  \bvol{86}~(1),  \pg{905860106}.

\bibitem[Kappel {\em et~al.\/}(2024)Kappel, Landreman \& Malhotra]{Kappel2024}
{\sc \au{Kappel, John}, \au{Landreman, Matt} \& \au{Malhotra, Dhairya}} \yr{2024}  \at{The magnetic gradient scale length explains why certain plasmas require close external magnetic coils}.  \jt{Plasma Physics and Controlled Fusion}  \bvol{66}~(2),  \pg{025018}.

\bibitem[Kingma \& Ba(2017)]{Adam17}
{\sc \au{Kingma, D.~P.} \& \au{Ba, J.}} \yr{2017}  \at{Adam: A method for stochastic optimization}.  \jt{arXiv 1412.6980} .

\bibitem[Landreman(2021)]{Landreman2021a}
{\sc \au{Landreman, M.}} \yr{2021}  \at{{Figures of merit for stellarators near the magnetic axis}}.  \jt{Journal of Plasma Physics}  \bvol{87}~(1),  \pg{905870112}.

\bibitem[Landreman(2022)]{Landreman_2022}
{\sc \au{Landreman, M.}} \yr{2022}  \at{Mapping the space of quasisymmetric stellarators using optimized near-axis expansion}.  \jt{Journal of Plasma Physics}  \bvol{88}~(6).

\bibitem[Landreman \& Jorge(2020)]{Landreman2020a}
{\sc \au{Landreman, M.} \& \au{Jorge, R.}} \yr{2020}  \at{{Magnetic well and Mercier stability of stellarators near the magnetic axis}}.  \jt{Journal of Plasma Physics}  \bvol{86}~(5),  \pg{905860510}.

\bibitem[Landreman {\em et~al.\/}(2021)Landreman, Medasani \& Zhu]{Landreman2021}
{\sc \au{Landreman, M.}, \au{Medasani, B.} \& \au{Zhu, C.}} \yr{2021}  \at{{Stellarator optimization for good magnetic surfaces at the same time as quasisymmetry}}.  \jt{Physics of Plasmas}  \bvol{28}~(9),  \pg{092505}.

\bibitem[Landreman \& Sengupta(2019)]{Landreman2019b}
{\sc \au{Landreman, M.} \& \au{Sengupta, W.}} \yr{2019}  \at{{Constructing stellarators with quasisymmetry to high order}}.  \jt{Journal of Plasma Physics}  \bvol{85}~(6),  \pg{815850601}.

\bibitem[Lu \& Lu(2020)]{Lu2020}
{\sc \au{Lu, Yulong} \& \au{Lu, Jianfeng}} \yr{2020} A universal approximation theorem of deep neural networks for expressing probability distributions,  \arxiv{arXiv: 2004.08867}.

\bibitem[McLachlan \& Basford(1988)]{McLachlan1988}
{\sc \au{McLachlan, G.~J.} \& \au{Basford, K.~E.}} \yr{1988} {\em Mixture models: Inference and applications to clustering\/}.  \publ{Marcel Dekker}.

\bibitem[McLachlan \& Peel(2004)]{mclachlan2004finite}
{\sc \au{McLachlan, G.~J.} \& \au{Peel, D.}} \yr{2004} {\em Finite Mixture Models\/}. {\em Wiley Series in Probability and Statistics\/} 1.  \publ{Wiley}.

\bibitem[Mercier(1964)]{Mercier1964}
{\sc \au{Mercier, C.}} \yr{1964}  \at{{Equilibrium and stability of a toroidal magnetohydrodynamic system in the neighbourhood of a magnetic axis}}.  \jt{Nuclear Fusion}  \bvol{4}~(3),  \pg{213}.

\bibitem[Murphy(2023)]{KMurphy2023}
{\sc \au{Murphy, K.}} \yr{2023} {\em Probabilistic Machine Learning: Advanced Topics\/}.  \publ{MIT Press}.

\bibitem[Nuhrenberg \& Zille(1988)]{Nuhrenberg1988}
{\sc \au{Nuhrenberg, J.} \& \au{Zille, R.}} \yr{1988}  \at{{Quasi-helically symmetric toroidal stellarators}}.  \jt{Physics Letters A}  \bvol{129}~(2),  \pg{113}.

\bibitem[Paul {\em et~al.\/}(2022)Paul, Bhattacharjee, Landreman, Alex, Velasco \& Nies]{Paul_2022}
{\sc \au{Paul, E.~J.}, \au{Bhattacharjee, A.}, \au{Landreman, M.}, \au{Alex, D.}, \au{Velasco, J.~L.} \& \au{Nies, R.}} \yr{2022}  \at{Energetic particle loss mechanisms in reactor-scale equilibria close to quasisymmetry}.  \jt{Nuclear Fusion}  \bvol{62}~(12),  \pg{126054}.

\bibitem[Solov'ev \& Shafranov(1970)]{Solovev1995}
{\sc \au{Solov'ev, L.~S.} \& \au{Shafranov, V.~D.}} \yr{1970}  \at{Plasma confinement in closed magnetic systems}.  \bt{In {\em Reviews of Plasma Physics\/} (ed. \ed{M.~A. Leontovich})}, ,  \vol{vol.~5},  \pg{pp. 1--247}.  \publ{Springer}.

\bibitem[Spitzer(1958)]{Spitzer1958}
{\sc \au{Spitzer, L.}} \yr{1958}  \at{{The stellarator concept}}.  \jt{Physics of Fluids}  \bvol{1}~(4),  \pg{253}.

\bibitem[Uzair \& Jamil(2020)]{Uzair2020}
{\sc \au{Uzair, Muhammad} \& \au{Jamil, Noreen}} \yr{2020} Effects of hidden layers on the efficiency of neural networks.  \bt{In {\em 2020 IEEE 23rd International Multitopic Conference (INMIC)\/}},  \pg{pp. 1--6}.

\end{thebibliography}

\end{document}